\documentclass[12pt]{article}

\pdfoutput=1

\usepackage[top=80pt,bottom=85pt,left=60pt,right=60pt]{geometry}
\usepackage{amssymb,amsmath,diagbox,multicol,multirow}
\usepackage{slashed}
\usepackage{graphicx,color,xcolor,subfigure}
\usepackage{float}
\usepackage{sectsty}
\usepackage{cite}

\makeatletter
\@addtoreset{equation}{section}
\makeatother

\newcommand{\beq}{\begin{equation}}
\newcommand{\eeq}{\end{equation}}
\newcommand{\bea}{\begin{eqnarray}}
\newcommand{\eea}{\end{eqnarray}}
\newcommand{\nn}{\nonumber}

\begin{document}

\begin{titlepage}

\begin{center}

\vskip .5in 
\noindent

{\Large \bf{${\cal N}=(1,1)$ supersymmetric AdS$_3$ in 10 dimensions}}

\bigskip\medskip
 Niall T. Macpherson$^{a,b}${\footnote{ntmacpher@gmail.com}}, Alessandro Tomasiello$^{c}${\footnote{alessandro.tomasiello@unimib.it}}\\

\bigskip\medskip
{\small }

${}^a$ Department of Physics, University of Oviedo,\\ Avda. Federico Garcia Lorca s/n, 33007 Oviedo,
Spain\\
\vskip 3mm
${}^b$ Departamento de F\'isica de Part\'iculas
Universidade de Santiago de Compostela\\
\vskip 1mm
and\\
\vskip 1mm
Instituto Galego de F\'isica de Altas Enerx\'ias (IGFAE)\\
E-15782 Santiago de Compostela, Spain\\
\vskip 3mm
${}^c$ Dipartimento di Matematica, Universit\`a di Milano--Bicocca,\\
Via Cozzi 55, 20126 Milano, Italy\\
\vskip 1mm
and\\
\vskip 1mm
INFN, sezione di Milano--Bicocca, Italy\\
\vskip 3mm

\vskip .9cm 
     	{\bf Abstract }\\
			~\\
			\end{center}
Warped AdS$_3$ solutions in 10 dimensional supergravity that preserve ${\cal N}=(1,1)$ supersymmetry are considered. Sufficient geometric conditions for their existence, and to stop the AdS$_3$ factor experiencing an enhancement to AdS$_4$, are presented. The internal space of such solutions decomposes as a foliation of M$_6$ over an interval where M$_6$ supports either an SU(3)- or SU(2)-structure. The former case is classified in terms of torsion classes and new solutions are found. 
\vskip .1in

\noindent

\noindent

\vfill
\eject

\end{titlepage}

\tableofcontents

\section{Introduction}
In recent years the AdS$_3$/CFT$_2$ correspondence has received increased attention. For the canonical maximally supersymmetric cases of the duality,\footnote{Recall that the notation ${\mathcal N}=(p,q)$ indicates that the two copies of $\mathrm{sl}(2,\mathbb{R})$ in the AdS$_3$ isometry algebra can be supersymmetrized independently.
For AdS$_3$ solution in 10 or 11 dimensions, $p+q\le 8$ \cite{Haupt:2018gap}. The ``small'' and ``large'' labels will be reviewed soon.} the D1--D5 \cite{Maldacena:1997re} and D1--D5+D1--D5 \cite{deBoer:1999gea} near horizons, preserving either small or large  ${\cal N}=(4,4)$ algebras, a great deal of progress has been made from both the string and CFT perspectives. Finding the CFT dual to \cite{deBoer:1999gea} had been a long standing open question which was finally answered in \cite{Eberhardt:2017fsi,Eberhardt:2017pty} --- the key contribution being an explicit computation of the (S-dual) Kaluza--Klein spectrum that had been lacking. It was realised how to quantise strings on the pure NS avatars of AdS$_3\times$S$^3\times \mathbb{T}^4$ and AdS$_3\times$S$^3\times$S$^3\times$S$^1$ in \cite{Gaberdiel:2018rqv,Eberhardt:2018ouy} and \cite{Eberhardt:2017fsi} respectively. These advancements have led to a deeper understanding of the AdS$_3$/CFT$_2$ correspondence, even leading to a proposed world-sheet derivation, for pure NS AdS$_3\times$S$^3\times \mathbb{T}^4$, in
\cite{Eberhardt:2019ywk}.\\
~\\
This progress motivates efforts to construct more general AdS$_3$ string vacua in $d=10,11$ that are dual to two dimensional CFTs. This is a wide and varied problem: as we will review shortly, a priori such vacua can come equipped with a wide array of possible superconformal algebras  \cite{Fradkin:1992bz,Beck:2017wpm}, but most of the allowed possibilities have not been realized so far. 
In particular, except for the ${\mathcal N}=(4,4)$ case, little systematic progress has been made studying geometries supporting both a left and right superconformal algebra. In this work we aim to provide some tools to redress this. Specifically we will derive a geometric framework for the construction of string vacua with ${\cal N}=(1,1)$ supersymmetry. This case has been relatively unexplored so far (one known example being in \cite{Gauntlett:2000ng}), but it may be a promising stepping stone towards more general ${\cal N}=(p,q)$ vacua.\\
~\\
As promised, we now give a short review of the general situation so far. Recall first that for some values of ${\mathcal N}$ there are several possible superalgebras, both for the L and R copy. For ${\mathcal N}=8$ there are four possibilities, with R-symmetry $\mathrm{so}(8)$, $\mathrm{spin}(7)$, $\mathrm{u}(4)$, $\mathrm{sp}(2)\oplus \mathrm{sp}(1)$. For ${\mathcal N}=7$ there are two possibilities: $\mathrm{so}(7)$ (large) and $\mathrm{g}_2$ (small). For ${\mathcal N}=6$, $\mathrm{so}(6)$ (large) and $\mathrm{u}(3)$ (small); and finally for ${\mathcal N}=4$, $\mathrm{so}(4)$ (large) and $\mathrm{so}(3)$ (small). In ${\mathcal N}=(4,4)$ one can then have three possibilities, and the (large,small) case is sometimes also called ``medium''. \\
~\\
At this point two maximal cases are completely classified. First the local form of all large ${\cal N}=(4,4)$ solutions was found across \cite{DHoker:2008lup,Estes:2012vm,Bachas:2013vza,Macpherson:2018mif}. Second, all ${\cal N}=(8,0)$ solutions were recently found in \cite{Legramandi:2020txf}  (see also \cite{Dibitetto:2018ftj} for the first such example and \cite{Deger:2019tem} for a 3d gauged supergravity perspective). The only true string vacua\footnote{\cite{Macpherson:2018mif} find more, but these are ${\cal N}=(4,0)$.} among these lie within the U-duality orbit of the D1--D5+D1--D5 near horizon; however  these works do find many interesting Janus-like solutions (putatively) dual to conformal defects within higher dimensional CFTs. The case of small ${\cal N}=(4,4)$ is more sparsely populated: we are only aware of the U-duality orbit of the D1--D5 near horizon 
\cite{Maldacena:1997re} and \cite{Dibitetto:2020bsh}, with a geometrical characterization given in \cite{Gauntlett:2007ph}. The status of the other cases with sixteen supercharges, ${\cal N}=(8-n,n)$, is unknown. Of the less supersymmetric examples:
There is an example of (small) ${\cal N}=(7,0)$ in \cite{Dibitetto:2018ftj}, while the status of ${\cal N}=(6,0)$ and ${\cal N}=(5,0)$ is unknown. Partial classifications results and some particular solutions for large and small ${\cal N}=(4,0)$ are given across  \cite{Macpherson:2018mif,Lozano:2015bra,Kelekci:2016uqv} and \cite{Maldacena:1997de,Kim:2007hv,OColgain:2010wlk,Couzens:2017way,Lozano:2019emq,Lozano:2019jza,Lozano:2019zvg,Lozano:2019ywa,Faedo:2020lyw,Faedo:2020nol,Zacarias:2021pfz,Couzens:2021veb} respectively. An example with ${\mathcal N}=(4,2)$ is given in \cite{Donos:2014eua}.
For ${\cal N}=(2,2)$, an incomplete picture is provided by \cite{Couzens:2021tnv,Eberhardt:2017uup,Gauntlett:2007ph}; explicit examples are in \cite{Maldacena:2000mw,Gauntlett:2001jj}. There are limited examples with ${\cal N}=(3,0)$ and ${\cal N}=(3,3)$ in \cite{Figueras:2007cn,Legramandi:2019xqd,Eberhardt:2018sce}. There is one known example of ${\mathcal N}=(2,1)$ solution in \cite[Sec.~2]{Gauntlett:2001jj}. A systematic study of ${\cal N}=(2,0)$ in IIB is provided by \cite{Kim:2005ez,Couzens:2017nnr,Couzens:2019iog}, and there exist several classifications of ${\cal N}=(1,0)$ \cite{Dibitetto:2018ftj,Martelli:2003ki,Tsimpis:2005kj,Passias:2020ubv,Passias:2019rga}; for the latter two, many explicit examples exist. We summarize this discussion in Table \ref{tab:ads3}.\\
\begin{table}[t]\label{tab:ads3}
\centering
\begin{tabular}{|c||*{8}{c}|}\hline
\diagbox{${\mathcal N}_\mathrm{R}$}{${\mathcal N}_\mathrm{L}$} & 1 & 2 & 3 & 4 & 5 & 6 & 7 & 8 \\
\hline\hline
0 &pcl\ &pcl\ &ex\ &pcl &?\ &?\  & ?(l) ex(s)\ & cl\\
1 &ex &ex &? &? &? &? &? & \\
2 & &ex &? &ex &? &? & & \\
3 & & &ex &? &? & & & \\
4 & & & &cl(l) ?(m) ex(s)  & & & & \\
\hline
\end{tabular}
\caption{\small Summary of known AdS$_3$ classification: ``cl'' means that a complete classification exists; ``pcl'' means partial classification; ``ex'' means that some examples are known; ``?'' means that no examples are known. As recalled in the text, some cases have a large (l) and small (s) version; ${\mathcal N}=4$ also has a medium (m) case. In ${\mathcal N}=8$ the classification exists for all four versions, while for ${\mathcal N}=6$ and 5 no example is known for either version.}
\end{table}
~\\
We will set up our study in terms of the pure spinor formalism (section \ref{sec:bispingen}, with a more detailed derivation in the appendices). As usual in this method, we reformulate supersymmetry in terms of equations in terms of pure forms and exterior algebra. 
We will find in general that the internal space M$_7$ is a fibration over an interval of a six-manifold M$_6$. For a more detailed analysis we restrict to the easiest case, where the internal supersymmetry parameters are characterized by an $\mathrm{SU}(3)$-structure $(J_2,\Omega_3)$ (section \ref{sec:param}). Here we identify several natural classes into which a solution must fall (sections \ref{sec:SU3IIA}, \ref{sec:SU3IIB}). \\
~\\
Two of these are particularly interesting: in IIA we find that M$_6$ has to be a half-flat manifold ($d J^2_2= d \mathrm{Re}  \Omega_3=0$), while in IIB we find an even less restrictive condition ($(d \mathrm{Re} \Omega_3)_{2,2}=0$). The former case is in particular not too surprising, as it already appears in a related context: G$_2$-holonomy spaces can be obtained as foliations of half-flat manifolds \cite{Hitchin:2001rw,Chiossi:2002tw}, a fact that was exploited for a proposal to extend mirror symmetry beyond the Calabi--Yau realm \cite{Gurrieri:2002wz,Tomasiello:2005bp}. There are further restrictions that relate these geometries to the other fields, but overall these are rather forgiving, and we are able to find several explicit examples, albeit mostly numerically (section \ref{sec:solutions}). These include cases where M$_6$ is a sphere fibration over three- or two-dimensional manifold with constant curvature; they have fully localized and back-reacted O8- and O5-plane sources, as we summarize in section \ref{sub:summary}.

\section{Geometric conditions for type II ${\cal N}=(1,1)$ AdS$_3$ vacua }\label{sec:bispingen}
Here we shall define an AdS$_3$ vacua in type II supergravity to be any solution with bosonic fields that respect the decomposition
\begin{equation}\label{eq:gendecomp}
\begin{split}
	ds^2&= e^{2A}ds^2(\text{AdS}_3)+ ds^2(\text{M}_7),\qquad H= e^{3A}h_0\text{vol}(\text{AdS}_3)+H_3,\\[2mm]
	F_{10}&= f_{\pm}+e^{3A}\text{vol}(\text{AdS}_3)\wedge \star_7\lambda(f_{\pm})	
\end{split}	
\end{equation}
where we take AdS$_3$ to have inverse radius $m$, the AdS$_3$ warp factor $e^{2A}$, dilaton $\Phi$ and magnetic components of the NS flux, and RR polyform $(H_3,f_{\pm})$ have support on M$_7$ only. Additionally $\lambda(X_n)= (-1)^{\left[\frac{n}{2}\right]}X_n$ and $\pm$ labels forms of even/odd degree, so  the upper signs should be taken in IIA the lower ones in IIB here and elsewhere.  For a true AdS$_3$ string vacuum M$_7$ should also be some bounded manifold, but this is not something we shall impose at this point.\\
~~\\
One can realise ${\cal N}=(1,1)$ supersymmetry in an AdS$_3$ solution by assuming that the associated 10 dimensional Majorana--Weyl Killing spinors decompose as
\beq
\epsilon_1= \zeta_+\otimes \theta_+\otimes \chi^1_++\zeta_-\otimes \theta_+\otimes \chi^1_-,\qquad \epsilon_2= \zeta_+\otimes \theta_{\mp}\otimes \chi^2_++\zeta_-\otimes \theta_{\mp}\otimes \chi^2_-
\eeq
where $\zeta_{\pm}$ are independent real Killing spinors on AdS$_3$ charged under SL(2)$_{\pm}\subset$ SO(2,2), which obey the Killing spinor equation
\beq
\nabla_{\mu}\zeta_{\pm}= \pm \frac{m}{2}\gamma^{(3)}_{\mu}\zeta_{\pm},
\eeq
 $\chi^{1,2}_{\pm}$ are 4 independent Majorana spinors on M$_7$ whose norms can be taken to satisfy
\beq\label{eq:norms}
|\chi^1_+|^2+|\chi^2_+|^2 =|\chi^1_-|^2+|\chi^2_-|^2 = 2e^{A},\qquad e^{A}(|\chi^1_+|^2-|\chi^2_+|^2) = -e^A(|\chi^1_-|^2-|\chi^2_-|^2)= c
\eeq
without loss of generality, for $c$ some arbitary constant. Finally $\theta_{\pm}$ are two dimensional vectors that are always required when one decomposes an even dimensional space in terms of two odd ones, as such they contain the 10 dimensional chirality labeled here by the $\pm$ subscript; the subscript on the internal $d=7$ spinors is just a label. In appendix \ref{sec:derivationsusy} we derive sufficient geometric conditions for an ${\cal N}=(1,1)$ solution to exist in terms of bi-linears formed from $\chi^{1,2}_{\pm}$, namely
\beq
\Psi^{st}_++ i \Psi^{st}_-= \chi^1_s\otimes \chi^{2\dag}_t,~~~  
s,t=\pm,\qquad \chi^1_s\otimes \chi^{2\dag}_t=\frac{1}{8}\sum_{n=1}^7\frac{1}{n!}\chi^{2\dag}_t\gamma_{a_n...a_1}\chi^1_s e^{a_1...a_n}
\eeq
where $\gamma_a$ are flat space 7 dimensional gamma matrices, $e^a$ are a corresponding vielbein and the subscript on $\Psi^{st}_{\pm}$ again refers to form degree. We shall now summarise these geometric conditions.
 First off, similar necessary and sufficient conditions for ${\cal N}=1$ AdS$_3$ vacua are given in \cite{Dibitetto:2018ftj,Passias:2020ubv}, and it should be no surprise that when a geometry supports both an ${\cal N}=(1,0)$ and an ${\cal N}=(0,1)$ set of these one has necessary and sufficient conditions for ${\cal N}=(1,1)$: these are   
\begin{align}
&e^{3A}h_0=-m c,\nn\\[2mm]
&d_{H_3}(e^{A-\Phi}\Psi^{++}_{\mp})\pm \frac{c}{16}f_{\pm}=0,\qquad d_{H_3}(e^{A-\Phi}\Psi^{--}_{\mp})\mp \frac{c}{16}f_{\pm}=0,\label{eq:bpsness}\\[2mm]
&(\Psi^{++}_{\mp},f_{\pm})_7= \mp \frac{1}{2}\big(m+\frac{1}{4}e^{-A} c h_0\big) e^{-\Phi}\text{vol}(\text{M}_7),~~~~ (\Psi^{--}_{\mp},f_{\pm})_7= \pm \frac{1}{2}\big(m+\frac{1}{4}e^{-A} c h_0\big) e^{-\Phi}\text{vol}(\text{M}_7)\nn\\[2mm]
&d_{H_3}(e^{2A-\Phi}\Psi^{++}_{\pm})\mp 2m e^{A-\Phi}\Psi^{++}_{\mp}=\frac{1}{8}e^{3A}\star_7\lambda(f_{\pm}),~~~~d_{H_3}(e^{2A-\Phi}\Psi^{--}_{\pm})\pm 2m e^{A-\Phi}\Psi^{--}_{\mp}=\frac{1}{8}e^{3A}\star_7\lambda(f_{\pm}),\nn
\end{align}
where $(X,Y)_7$ is the 7 dimensional  Mukai pairing, namely  $(X,Y)_7=X\wedge \lambda(Y)\big\lvert_7$.  These conditions imply the Bianchi identity of the electric component of the RR flux, which when $dH_3=0$ is
\beq
d(e^{3A}\star_7\lambda(f_{\pm}))+ e^{3A}h_0 f_{\pm}=0
\eeq
and when $c \neq 0$ imply the source free magnetic flux Bianchi identity
\beq
d_{H_3}f_{\pm}=0.
\eeq
In addition to \eqref{eq:bpsness} it is possible to derive several more conditions involving bi-linears that mix the ${\cal N}=(1,0)$ and ${\cal N}=(0,1)$ sub-sectors; these are implied by what we present so far, but are very useful for constraining the system. First off it is possible to show that an ${\cal N}=(1,1)$ AdS$_3$ solution will experience an  enhancement to AdS$_4$ at all regular points of the internal space unless
\beq\label{eq:notAdS4cond}
\chi^{1\dag}_-\chi^1_++\chi^{2\dag}_-\chi^2_+=0,\qquad \chi^{1\dag}_+\gamma_a\chi^1_-\mp \chi^{2\dag}_+\gamma_a\chi^2_-=0;
\eeq
so, if one is interested in genuine AdS$_3$ solutions, these conditions should be satisfied.\footnote{The argument leading to this claim is local, but may hold in general. The metric is always locally AdS$_4$ when \eqref{eq:notAdS4cond} is not imposed, but  we have not ruled out the possibility of fluxes that contain terms that break the isometries of AdS$_4$ when their Bianchi identities are violated, either by partially localised or completely smeared sources.} Under these restrictions one can show that several other conditions should be satisfied: these additionally involve 
\beq
g=\chi^{1\dag}_-\chi^1_+-\chi^{2\dag}_-\chi^2_+,\qquad  \tilde{\xi}=-i(\chi^{1\dag}_+\gamma_a\chi^1_-\pm \chi^{2\dag}_+\gamma_a\chi^2_-) e^a,
\eeq
neither of which can vanish globally, otherwise ${\cal N}=(1,1)$ cannot be realised. The conditions are
\begin{align}
&d(e^{A}g)+m  \tilde{\xi}=0,\nn\\[2mm]
&d_{H_3}(e^{2A-\Phi}(\Psi^{+-}_{\pm}+\Psi^{-+}_{\pm}))=0,\nn\\[2mm]
&d_{H_3}(e^{-\Phi}(\Psi^{+-}_{\pm}-\Psi^{-+}_{\pm}))=\frac{1}{8}\tilde{\xi}\wedge f_{\pm},\label{eq:extrabps}\\[2mm]
&d_{H_3}(e^{A-\Phi}(\Psi^{+-}_{\mp}+\Psi^{-+}_{\mp}))\pm m e^{-\Phi}(\Psi^{+-}_{\pm}-\Psi^{-+}_{\pm})=\mp\frac{1}{8}e^{A}g f_{\pm},\nn,\\[2mm]
&d_{H_3}(e^{3A-\Phi}(\Psi^{+-}_{\mp}-\Psi^{-+}_{\mp}))\pm e^{3A-\Phi}h_0(\Psi^{+-}_{\pm}-\Psi^{-+}_{\pm})\pm 3m e^{2A-\Phi}(\Psi^{+-}_{\pm}+\Psi^{-+}_{\pm})= \pm \frac{1}{8}e^{3A}\tilde{\xi}\wedge \star_7\lambda(f_{\pm})\nn
\end{align}
from which it follows that the Bianchi identity of the magnetic flux is in fact implied in general for ${\cal N}=(1,1)$ AdS$_3$ vacua, when that of $H$ is imposed. Much of \eqref{eq:bpsness}, \eqref{eq:extrabps} can be shown to be redundant using some identities that the 7d bi-linears must obey, \eqref{eq:identities1}--\eqref{eq:identities8}. Upon tuning the internal spinors such that \eqref{eq:norms} and \eqref{eq:notAdS4cond} are satisfied, a sufficient system for ${\cal N}=(1,1)$ supersymmetry is given by\footnote{It turns out that when $c=0$ and the internal manifold supports an SU(3)-structure, the main focus of this work, \eqref{eq:reducedBPS6} and \eqref{eq:reducedBPS8} are actually implied. We suspect that this holds true in general but have not proved this.}
\begin{subequations}
\begin{align}
&e^{3A}h_0=-m c,\label{eq:reducedBPS1}\\[2mm]
&d(e^{A}g)+m  \tilde{\xi}=0,\label{eq:reducedBPS2}\\[2mm]
&d_{H_3}(e^{A-\Phi}(\Psi^{++}_{\mp}+\Psi^{--}_{\mp}))=0,\label{eq:reducedBPS3}\\[2mm]
&d_{H_3}(e^{2A-\Phi}(\Psi^{++}_{\pm}-\Psi^{--}_{\pm}))\mp 2m e^{A-\Phi}(\Psi^{++}_{\mp}-\Psi^{--}_{\mp})=0,\label{eq:reducedBPS4}\\[2mm]
&d_{H_3}(e^{A-\Phi}(\Psi^{++}_{\mp}-\Psi^{--}_{\mp}))\pm \frac{c}{8}f_{\pm}=0,\label{eq:reducedBPS5}\\[2mm]
&d_{H_3}(e^{2A-\Phi}(\Psi^{++}_{\pm}+\Psi^{--}_{\pm}))\mp 2m e^{A-\Phi}(\Psi^{++}_{\mp}+\Psi^{--}_{\mp})=\frac{1}{4}e^{3A}\star_7\lambda(f_{\pm}),\label{eq:reducedBPS6}\\[2mm]
&d_{H_3}(e^{A-\Phi}(\Psi^{+-}_{\mp}+\Psi^{-+}_{\mp}))\pm m e^{-\Phi}(\Psi^{+-}_{\pm}-\Psi^{-+}_{\pm})=\mp\frac{1}{8}e^{A}g f_{\pm},\label{eq:reducedBPS7}\\[2mm]
&(\Psi^{++}_{\mp}-\Psi^{--}_{\mp},f_{\pm})_7= \pm \big(m+\frac{1}{4}e^{-A} c h_0\big) e^{-\Phi}\text{vol}(\text{M}_7)\label{eq:reducedBPS8}.
\end{align}
\end{subequations}
To actually have a solution one must also solve the supergravity equations of motion, however well known integrability arguments tell us that these are implied for AdS$_3$ by supersymmetry and the Bianchi identities of $f_{\pm}$ and $H_3$. For ${\cal N}=(1,1)$ specifically this amounts to just imposing
\beq
dH_3=0
\eeq 
in regular regions of the internal space, with possible $\delta$-function sources elsewhere.  The remaining equations of motion then follow. Note also that a non trivial Romans mass in IIA requires $c=h_0=0$ due to \eqref{eq:reducedBPS4} while in IIB one can always assume one is in an SL(2,$\mathbb{R}$) duality frame where $h_0=c=0$ holds. Thus in the rest of the paper we will simply fix $c=0$, with the understanding that the only IIB solutions outside this limit are S-dual to what we do find, and the only IIA solutions belong to classes better studied from an M-theory perspective.\footnote{In other words, all such IIA solutions can be lifted to M-theory where they will have a U(1) flavour isometry. Such solutions may just be special cases of broader classes without this U(1), but one is blind to that in IIA.}\\
~~\\
In the next section we shall proceed to parametrise the bi-linears appearing in \eqref{eq:reducedBPS1}--\eqref{eq:reducedBPS7} in the $c=0$ limit.

\section{Parametrising the bi-linears}\label{sec:param}

As argued in the previous section, to end up with a solution that is not locally AdS$_4$ we should solve \eqref{eq:notAdS4cond}. We will assume from now on that $c=0$, or in other words that the internal spinors have equal norm; this is often necessary in higher dimensions. Now \eqref{eq:reducedBPS1} is solved by $h_0=0$; this means that without loss of generality we can decompose our four Majorana spinors in a basis of two real unit norm vectors $(V,\hat {V})$, three real functions $(a,b,\alpha)$ and a single unit norm spinor $\chi$ as
\begin{align}
\chi^1_+&= e^{\frac{A}{2}}\chi,& \chi^1_-&= e^{\frac{A}{2}}(\cos\alpha+ i \sin\alpha V)\chi,\\[2mm]
 \chi^2_+&= e^{\frac{A}{2}}(a+i b  V)\chi,& \chi^2_-&=-e^{\frac{A}{2}}(\cos\alpha\mp i \sin\alpha V)(a+i b  \hat {V})\chi,\nn
\end{align}
where $a^2+b^2=1$, which immediately leads to 
\beq
g= 2e^{A}\cos\alpha,\qquad \tilde{\xi}=2 e^{A}\sin\alpha  V
\eeq
where neither $\sin\alpha$ nor $\cos\alpha$ can be set to zero globally. This means that \eqref{eq:reducedBPS2} becomes
\beq
d(e^{2A}\cos\alpha)+ m e^{A}\sin\alpha V=0\label{eq:vgen},
\eeq
which implies that locally M$_7$ decomposes as a warped product of an interval spanned by $V$ and some six-manifold M$_6$. Locally one can solve \eqref{eq:vgen} in terms of a function $h$ of a local coordinate $y$ as
\beq\label{eq:localV}
e^{2A}\cos\alpha= h,~~~~  V=- \frac{h'}{m e^{A}\sqrt{1-e^{-4A}h^2}} dy 
\eeq
where $h$ essentially parametrises diffeomorphism invariance in $y$, and $h'=\partial_y h$. The metric then decomposes as
\beq \label{eq:M7-M6}
ds^2(\text{M}_7)=\frac{( h')^2}{m^2 e^{2A}(1-e^{-4A}h^2)} dy^2+ ds^2(\text{M}_6).
\eeq
The specific form M$_6$ can take depends on precisely how the rest of the supersymmetry conditions are solved, in general its metric will depend on $y$.\\
~~\\
In general we can take the spinor $\chi$ to be such that
\beq
\chi\otimes \chi^{\dag}=\Psi^{(\text{G}_2)}_++ i \Psi^{(\text{G}_2)}_-= \frac{1}{8}(1-i\Phi_3-\star_7 \Phi_3+ i\text{vol}_7),\qquad \Phi_3\wedge \star_7\Phi_3=7 \text{vol}_7,
\eeq
where $\Phi_3$ is the 3 form associated to the  G$_2$-structure that a single  $d=7$ Majorana spinor supports. The actual G-structure supported by an ${\cal N}=(1,1)$ solution will depend on how $\hat {V}$ is aligned, if it is parallel to $V$ then the solution will support an SU(3)-structure on $\text{M}_6$, otherwise (assuming $b\neq 0$) it will support an SU(2)-structure there. In general we can decompose the G$_2$-structure in terms of an SU(3)-structure  as
\beq
\Phi_3=  V\wedge J_2-\text{Im}\Omega_3,\qquad \star_7\Phi_3= \frac{1}{2}J_2\wedge J_2-V\wedge\text{Re}\Omega_3
\eeq
where $(J_2,\Omega_3)$ are the associated $(1,1)$ and $(3,0)$ forms that are orthogonal to $V$. When $b$ is non trivial we can further decompose 
\beq
\hat V= \tilde{\kappa}_{\|} V+ \tilde{\kappa}_{\perp} U_1,\qquad  U_1\lrcorner V=0,\qquad  U_1\lrcorner U_1=1,\qquad \tilde{\kappa}_{\|}^2+\tilde{\kappa}_{\perp}^2=1,
\eeq
and then the SU(3)-structure decomposes in terms of an SU(2)-structure orthogonal to $V$ as
\beq
J_2= j_2 +U_1\wedge U_2,\qquad \Omega_3=(U_1+i U_2)\wedge \tilde{\omega}_2 ,\qquad U_1\lrcorner U_2=0,\qquad U_2\lrcorner U_2=1.
\eeq
where $(j_2,\tilde{\omega}_2)$ define a 4d SU(2)-structure orthogonal to $(U_1,U_2)$ which, when  $b \kappa_{\|} \neq 0$, one can take to be two components of the 7 dimensional vielbein without loss of generality. One can then show that the bi-linears of the previous section decompose in terms of two bi-linears on M$_6$ $\Psi^{(6)}_{\pm}$ as
\begin{align}
\Psi^{++}_{+}&= e^{A}\text{Re}\bigg(\Psi^{(6)}_++V\wedge \Psi^{(6)}_-\bigg),\qquad \Psi^{++}_{-}= e^{A}\text{Im}\bigg(\Psi^{(6)}_-+V\wedge \Psi^{(6)}_+\bigg),\nn\\[2mm]
\Psi^{-+}_{+}&=e^{A}\text{Re}\bigg(e^{i\alpha}(\Psi^{(6)}_++V\wedge \Psi^{(6)}_-)\bigg),\qquad \Psi^{-+}_{-}=e^{A}\text{Im}\bigg(e^{i\alpha}(\Psi^{(6)}_-+V\wedge \Psi^{(6)}_+)\bigg),\label{eq:bispinors1}\\[2mm]
\Psi^{+-}_{+}&=-e^{A}\text{Re}\bigg(e^{\pm i\alpha}\Psi^{(6)}_++e^{\mp i\alpha}V\wedge\Psi^{(6)}_- \bigg),\qquad \Psi^{+-}_{-}=-e^{A}\text{Im}\bigg(e^{\mp i\alpha}\Psi^{(6)}_-+e^{\pm i\alpha}V\wedge\Psi^{(6)}_+\bigg), \nonumber \\[2mm]
\Psi^{--}_{+}&=-e^{A}\text{Re}\bigg(e^{2i\alpha_{\pm}}\Psi^{(6)}_++e^{2i\alpha_{\mp}}V\wedge\Psi^{(6)}_-\bigg),~~~\Psi^{--}_{-}=-e^{A}\text{Im}\bigg(e^{2i\alpha_{\mp}}\Psi^{(6)}_-+e^{2i\alpha_{\pm}}V\wedge\Psi^{(6)}_+ \bigg)\nn,
\end{align}
where $\alpha_+=\alpha$ and $\alpha_-=0$, which parameterise the difference between IIA and IIB bi-linears. The 6d bi-linears themselves may be expressed in general as
\beq
\Psi^{(6)}_+= \frac{e^{i\beta}}{8}(\kappa_{\|}e^{-i j_2}+\kappa_{\perp}\omega_2)\wedge e^{\frac{1}{2}U\wedge \overline{U}},\qquad  \Psi^{(6)}_-=\frac{1}{8}U\wedge (\kappa_{\|}\omega_2- \kappa_{\perp} e^{-i j_2}),\label{eq:bispinors2}
\eeq
where we have simplified the parametrisation by introducing 
\beq
(a+i b \tilde{\kappa}_{\|})= e^{i \beta}\kappa_{\|},\qquad  b\tilde{\kappa}_{\perp}=\kappa_{\perp},\qquad \tilde{\omega}_2= e^{i\beta}\omega_2,~~~\kappa_{\|}^2+\kappa_{\perp}^2=1,\qquad  U= U_1+i U_2.
\eeq
which are standard bi-linears for a 6d  SU(2)-structure.\\
~~\\
In this work we shall focus on the SU(3)-structure limit, $(\kappa_{\|},\kappa_{\perp})=(1,0)$ where the  bi-linears reduce to
\beq
\Psi^{(6)}_+= \frac{e^{i\beta}}{8}e^{-i J_2},\qquad \Psi^{(6)}_-=\frac{1}{8}\Omega_3,
\eeq
leaving the general case for future work. To this end it is helpful to review some features of SU(3)-structures in 7 dimensions. First we should make our conventions clear. We have taken $(J_2,\Omega_3)$ to be defined such that 
\begin{equation}
\begin{split}
	&J_2\wedge\Omega_3=0,\qquad J_2\wedge J_2\wedge J_2=\frac{3i}{4}\Omega_3\wedge \overline{\Omega}_3,\label{eqSU3conds}\\[2mm]
	&\star_7 1=\text{vol}_7=\frac{1}{3!} V\wedge J_2\wedge J_2\wedge J_2,\qquad  \star_7\Omega_3= i V\wedge \Omega_3.
\end{split}	
\end{equation}
This fixes the torsion classes of the SU(3)-structure to be \cite{DallAgata:2003txk} 
\begin{align}\label{eq:torsions}
dV&=RJ_2+T_1+\text{Re}(\iota_{\overline{V}_1}\Omega_3)+V\wedge W_0,\nn\\[2mm]
dJ_2&=\frac{3}{2}\text{Im}(\overline{W}_1\Omega_3)+W_3+W_4\wedge J_2+V\wedge\bigg(\frac{2}{3}\text{Re}E J_2+T_2+\text{Re}(\iota_{\overline{V}_2}\Omega_3)\bigg),\\[2mm]
d\Omega_3&=W_1 J_2\wedge J_2+W_2\wedge J_2+ \overline{W}_5\wedge \Omega_3+ V\wedge\bigg(E\Omega_3-2 V_2\wedge J+S\bigg). \nonumber
\end{align}
The part without $V$ in $dJ_2$ and $d \Omega_3$ reduces to the more familiar definition of torsion classes for SU(3)-structures in $d=6$.
$(E,R)$ are complex and real functions respectively, $(V_{1,2},W_5)$ are (1,0)-forms, $(W_0,W_4)$ real one-forms, $(W_4,T_{1,2})$ are respectively complex and real primitive\footnote{See \cite{Passias:2019rga} for a very good review of what this means in this context.} (1,1)-forms, $S$ is a primitive $(2,1)$ form and $W_3$ is the real part of such a form. These make up irreducible representations of SU(3), the details of which are immaterial for our purposes; what matters is that they must obey the identities
\begin{align}
&(S,W_3)\wedge J_2=(S,W_2,V_{1,2},W_5)\wedge\Omega_3=(W_4,T_{1,2})\wedge J_2\wedge J_2=0,\nn\\[2mm]
&\star_7S= -i V\wedge S,\qquad  \star_7 (W_4,T_{1,2})= -V\wedge J_2\wedge (W_4,T_{1,2}),\label{eq:id}\\[2mm] 
&\star_7(V_{1,2},W_5)=\frac{i}{2} V\wedge J_2\wedge J_2\wedge (V_{1,2},W_5),\nonumber
\end{align}
where $W_4$ behaves under Hodge duality as  $\text{Re}S$ does.  These facts will enable us to establish what manifolds can support ${\cal N}=(1,1)$ supersymmetry in terms of an SU(3)-structure in the following sections. Solving the supersymmetry constraints will set some of $W_i$ to zero, and will consequently also lead to constraints on the $y$-dependent M$_6$ in the decomposition (\ref{eq:M7-M6}). As we recall in table \ref{table:one}, many manifolds where some of the $W_i$ vanish have been studied and have known names.\footnote{Some of the names are rather common, some less so. We are mostly following notation in \cite{gray-hervella}, which however was concerned with U(3)- (rather than SU(3)-) structures, where $W_5$ is not defined. Following \cite{Grana:2005jc}, we have set $W_5=0$ except for the standard symplectic, complex and K\"ahler cases. The nearly and almost K\"ahler cases might perhaps have been called nearly and almost Calabi--Yau.} 
~\\
Finally we should stress that for all our ${\cal N}=(1,1)$ classes we necessarily have
\beq
R= V_1= T_1=0,
\eeq
with $W_0$ fixed universally due to \eqref{eq:vgen}, which holds in both IIA and IIB --- we shall thus not comment further on these torsion classes.\\
~\\
We begin our analysis in type IIA in the next section.
\begin{table}[h!]
\centering
\begin{tabular}{||c | c ||} 
 \hline
M$_6$ & Vanishing torsion classes\\ [0.5ex] 
 \hline\hline
Complex & $W_1=W_2=0$\\
Symplectic& $W_1=W_3=W_4=0$\\
Half-flat& $\text{Re}W_1=\text{Re}W_2=W_4=W_5=0$\\
Special Hermitian& $W_1=W_2=W_4=W_5=0$\\
Nearly K\"ahler& $W_2=W_3=W_4=W_5=0$\\
Almost K\"ahler& $W_1=W_3=W_4=W_5=0$\\
K\"ahler&  $W_1=W_2=W_3=W_4=0$\\
Calabi--Yau&  $W_1=W_2=W_3=W_4=W_5=0$\\
 \hline
\end{tabular}
\caption{\small Some notable SU(3)-structure six-manifolds in terms of vanishing torsion classes. Additionally a manifold may be conformally one of these when $(W_4,W_5)$ are appropriately tuned.}
\label{table:one}
\end{table}

\section{SU(3)-structure in IIA}\label{sec:SU3IIA}
In this section we shall classify the possible solutions in IIA that support an SU(3)-structure. As we shall see there are two classes where the 7d internal space decomposes as a foliation over an interval whose leaves are either an almost K\"ahler or half-flat six-manifold.\\
~~\\
To begin with we find it helpful to redefine the bi-linears of the previous section slightly by mapping
\beq
\beta\to \beta-\alpha,~~~\Omega_3\to e^{-i\frac{\pi}{2}}\Omega_3.
\eeq
It is then possible to show that the necessary conditions \eqref{eq:reducedBPS2}--\eqref{eq:reducedBPS4} are implied in general by
\begin{subequations}
\begin{align}
&d(e^{-A-\Phi}\frac{\cos\beta}{\cos\alpha })=d(e^{A-\Phi}\sin\beta \cot\alpha)\wedge V= 0,\label{IIAgenbps2}\\[2mm]
&d(e^{3A-\Phi}\cos\alpha \sin\beta J_2)-e^{3A-\Phi} \cos\alpha \cos\beta H_3+2 m e^{2A-\Phi} \sin\alpha\sin\beta V\wedge J_2=0,\label{IIAgenbps3}\\[2mm]
&d(e^{2A-\Phi}(\text{Re}\Omega_3+ \cos\alpha\cos\beta V\wedge J_2)+e^{2A-\Phi}\cos\alpha \sin\beta H_3 \wedge V=0,\label{IIAgenbps4}\\[2mm]
&\cos\alpha\sin\beta J_2\wedge J_2\wedge dJ_2-(\cos\alpha \cos\beta J_2\wedge J_2-3 V\wedge \text{Im}\Omega_3)\wedge H_3=0,\label{IIAgenbps5}\\[2mm]
&(\sin\alpha J_2\wedge d\text{Re}\Omega_3-\cos\beta d\alpha\wedge V\wedge J_2\wedge J_2= \sin\alpha H_3\wedge \text{Re}\Omega_3+\sin\beta d\alpha V\wedge J_2\wedge J_2=0,\label{IIAgenbps6}\\[2mm]
&d(e^{3A-\Phi}(\cos\alpha\cos\beta J_2\wedge J_2-2 V\wedge \text{Im}\Omega_3)+2 e^{3A-\Phi}\cos\alpha\sin\beta H_3\wedge J_2\nn\\[2mm]
&+2m e^{2A-\Phi}\sin\alpha\cos\beta V\wedge J_2\wedge J_2=0\label{IIAgenbps7}.
\end{align}
\end{subequations}
From \eqref{IIAgenbps3} we see that solutions with $\cos\beta=0$ appear to be distinct from those with generic values of $\beta$: namely this constraint fixes $H_3$ uniquely unless $\cos\beta=0$, in which case it is only more weakly constrained by the remaining conditions. This is a strong indication that $\cos\beta=0$ is a physically distinct class to generic $\beta$. It thus makes sense to study these two scenarios separately; we shall continue our analysis for the $\cos\beta=0$ case in the next section, and consider the more generic case in section \ref{sec:IIAbetagen}. 

\subsection{Class I: Restricted almost K\"ahler foliation $F_0=0$}\label{sec:IIAbetafixed}
This class is defined by fixing $\cos\beta=0$, and in fact we can solve this without loss of generality as
\beq
\beta=\frac{\pi}{2}.
\eeq 
Clearly the conditions \eqref{IIAgenbps2}--\eqref{IIAgenbps7} truncate a fair bit under this restriction.  First the conditions that remain non trivial in \eqref{IIAgenbps2} can be solved in terms of an arbitrary function $k(y)$ as
\beq
\frac{e^{-A-\Phi}h}{\sqrt{1- e^{-4A}h^2}} = k,
\eeq
where we use \eqref{eq:localV} to substitute for $\alpha$ here and elsewhere. To proceed it is helpful to decompose the NS flux as
\beq
H_3= dy\wedge H_2+ \tilde{H}_3,\qquad dy\lrcorner H_2=dy\lrcorner \tilde{H}_3=0,
\eeq
and to substitute for $(dJ_2,d\Omega_3)$ with their respective torsion classes  \eqref{eq:torsions} in \eqref{IIAgenbps3}--\eqref{IIAgenbps7}. A careful analysis reveals that $A=A(y)$ and that most of the torsion classes must vanish, specifically one has
\beq\label{eq:torsionsIIAcaseI} 
V_2=\text{Im}E=T_1=T_2=W_1=\text{Re}W_2=W_3=W_4=W_5=0.
\eeq
Comparing this to table \ref{table:one} tells us that M$_6$ is a $y$-dependent almost K\"ahler manifold, further restricted by the requirement that $\mathrm{Re}W_2=0$. These requirements are rather stringent, and in practice we will satisfy them in later sections only by taking conformal Calabi--Yau spaces.

The torsion classes that remain non trivial are
\beq\label{eq:torsionsIIAcaseI0}
\text{Re}E= 3m \frac{\partial_y(e^{A}h)}{h h' \sqrt{1- e^{-4A}h^2}},
\eeq
and $(S,\text{Im}W_2)$, which appear in the NS flux and get fixed as
\beq
H_3= \frac{e^{A} h'}{mh \sqrt{1- e^{-4A}h^2}}dy\wedge \text{Im}W_2+  \frac{e^{2A}}{h}\text{Re}S+ \frac{m\partial_y(e^{A} h)}{e^{2A} h' \sqrt{1- e^{-4A}h^2}}\text{Re}\Omega_3.
\eeq
Finally the functions $h,k$ are constrained to obey
\beq\label{eq:bpsIIAcaseI}
\partial_y\log\left(\frac{k}{h}\right)= \frac{3}{1- e^{-4A} h^2} \partial_y\log h,
\eeq
at which point \eqref{IIAgenbps2}--\eqref{IIAgenbps7} are implied. To make further progress one needs to extract the magnetic components of the RR flux from \eqref{eq:reducedBPS7}, which yields
\begin{align}
f_0&=0,~~~~f_2=\frac{m ke^{2A}}{h^2}\sqrt{1- e^{-4A}h^2} J_2,~~~~f_6=\frac{m}{3!}\frac{k\sqrt{1- e^{-4A}h^2} \partial_y(e^{4A}h^3)}{e^{2A} h^4h'} J_2\wedge J_2\wedge J_2,\\[2mm]
f_4&=\frac{e^{3A}k}{h^2}(1- e^{-4A}h^2) \text{Im}W_2\wedge J_2+  \frac{e^{2A}k h'}{m h^2}\sqrt{1- e^{-4A}h^2}dy\wedge \text{Im}S+\frac{k}{h}\partial_y(e^{-A}h)dy\wedge \text{Im}\Omega_3\nn.
\end{align}
These definitions of the flux actually imply \eqref{eq:reducedBPS6} and \eqref{eq:reducedBPS8}, so all of the supersymmetry constraints are implied by \eqref{eq:torsionsIIAcaseI}, \eqref{eq:torsionsIIAcaseI0} and \eqref{eq:bpsIIAcaseI}. The Bianchi identities of the RR fluxes and the remaining supergravity equations of motion now follow from the Bianchi identity of $H_3$, but this is not implied for this class so one additionally needs to impose it.

\subsection{Class II: Half-flat foliation $F_0 \neq 0$}\label{sec:IIAbetagen}
For this class we assume $\cos\beta\neq 0$, which means that we can divide by this quantity and further simplify \eqref{IIAgenbps2}--\eqref{IIAgenbps7}.  Similarly to the previous class one finds that
\beq
A= A(y),\qquad \beta= \beta(y),
\eeq
while this time the NS flux gets fixed such that it automatically solves its Bianchi identity as
\beq
H_3= d(\tan\beta J_2).
\eeq
The rest of \eqref{IIAgenbps2}--\eqref{IIAgenbps7} impose the following on the functions of the ansatz:
\beq\label{eq:scalarconstraints}
e^{A-\Phi}\cos\beta =c_1 h,\qquad \beta'= \sin(2\beta)\frac{\partial_y\log(e^{A}h)}{1- e^{-4A}h^2}
\eeq
where $c_1$ is an integration constant and fix the torsion classes. The following torsion classes must vanish:
\begin{align}\label{eq:IIAC2torsions1}
V_2=\text{Im}E=\text{Re}W_1=\text{Re}W_2=W_4= W_5=0,
\end{align}
making M$_6$ a $y$-dependent half-flat manifold. As we remarked in the introduction, the occurrence of such spaces is not surprising, because of the role they played in foliated G$_2$-manifolds \cite{Hitchin:2001rw,Chiossi:2002tw}. As we will see in later sections, it is quite easy to provide explicit examples.

The non trivial torsion classes are expressed in terms of $(T_2,S)$, which remain arbitrary, and the functions of the ansatz as
\begin{equation}\label{eq:IIAC2torsions2}
\begin{split}
	\text{Im}W_1&=-\frac{2m}{3}\frac{\cos\beta \partial_y(e^Ah)}{e^{2A} h'\sqrt{1- e^{-4A}h^2}},\qquad \text{Im}W_2=-\frac{h}{e^{2A}\cos\beta}T_2,\\[2mm]
	W_3&=\frac{e^{2A}\cos\beta}{h} \text{Re}S,~~~~ \text{Re} E=\frac{m(2-\cos(2\beta))\partial_y(e^{A}h)}{hh'\sqrt{1- e^{-4A}h^2}}
\end{split}	
\end{equation}
From  \eqref{eq:reducedBPS7} we extract the magnetic components of the RR fluxes:
\begin{align}
f_0=& F_0= c_1 m,\qquad f_2= c_1 m \tan\beta J_2,\qquad f_6= \frac{cm}{3!}\tan\beta \frac{\partial_y(e^{4A}h^3)}{e^{4A}h^2 h'}J_2\wedge J_2\wedge J_2, \nonumber\\[2mm]
f_4=& \frac{c h \sqrt{1- e^{-4A}h^2}}{e^A \cos^2\beta}J_2\wedge T_2+ \frac{c h'}{m\cos\beta}dy\wedge \text{Im}S\\
&+\frac{c h \partial_y(e^{-A}h)}{e^{2A}\cos\beta \sqrt{1- e^{-4A}h^2}}dy\wedge \text{Im}\Omega_3+\frac{c m}{3!}\frac{\partial_y(e^{4A}h)}{e^{4A} h'}J_2\wedge J_2;\nn
\end{align}
these again imply \eqref{eq:reducedBPS6} and \eqref{eq:reducedBPS8}. Thus as the Bianchi identities of the RR and NS fluxes are implied for this class one just needs to find half-flat manifolds consistent with \eqref{eq:IIAC2torsions1} and \eqref{eq:IIAC2torsions2} and solve \eqref{eq:scalarconstraints} to have a solution. We note that $\beta=0$, the trival solution to \eqref{eq:scalarconstraints}, is a potentially interesting limit as the flux reduces to just $(f_0,f_4)$. This suggests a link to the AdS$_6$ compactification on hyperbolic spaces $\Sigma_3$ appearing in \cite{Nunez:2001pt}; we will make this link more precise in section \ref{ssub:S3-IIA}.
~\\
This concludes our analysis of the geometries and fluxes of ${\cal N}=(1,1)$ solution preserving an SU(3)-structure in IIA. In the next section we shall carry out a similar analysis in IIB, before giving some new solutions within these classes in section \ref{sec:solutions}.

\section{SU(3)-structure in IIB}\label{sec:SU3IIB}
In this section we will classify SU(3)-structure solutions in IIB. As with IIA, the internal 7-manifold will decompose as a foliation of M$_6$ over an interval. Again there are two cases: one with M$_6$ again an almost K\"ahler manifold, the other is a quite broad class with $\text{Re}W_1=\text{Re}W_2=0$, but that is not merely conformally half-flat in general.\\
~~\\
As with IIA it is helpful to redefine the SU(3)-structure slightly, this time by just modifying the three-form as
\beq
\Omega_3\to -i e^{-i\alpha}\Omega_3.
\eeq
With some work it is then possible to show that the conditions \eqref{eq:reducedBPS2}--\eqref{eq:reducedBPS4} are implied by
\begin{subequations}
\begin{align}
&d(e^{2A-\Phi}\cos\beta)= d\left(e^{-\Phi}\frac{\sin\beta}{\sin(2\alpha)}\right)\wedge V=0,\label{eq:bpsIIB2}\\
&d(e^{2A-\Phi}\sin\beta J_2)-e^{2A-\Phi}\cos\beta H_3=0,\label{eq:bpsIIB3}\\[2mm]
&d(e^{3A-\Phi}(\cos\alpha \text{Re}\Omega_3+ \cos\beta J_2\wedge V))+e^{3A-\Phi}\sin\beta H_3\wedge V+2 me^{2A-\Phi} \sin\alpha V\wedge \text{Re}\Omega_3=0,\label{eq:bpsIIB4}\\[2mm]
&d(e^{2A-\Phi}(\cos\beta J_2\wedge J_2-2\cos\alpha V\wedge \text{Im}\Omega_3)+2 e^{2A-\Phi}\sin\beta J_2\wedge H_3=0\label{eq:bpsIIB5}\\[2mm]
&\sin\beta d\alpha\wedge V\wedge J_2\wedge J_2+\sin\alpha H_3\wedge \text{Re}\Omega_3=0,\label{eq:bpsIIB6}\\[2mm]
&d(e^{2A-\Phi}\sin\beta)\wedge J_2\wedge J_2\wedge J_2-3 e^{2A-\Phi}\cos\alpha V\wedge \text{Im}\Omega_3\wedge H_3=0,\label{eq:bpsIIB7}
\end{align}
\end{subequations}
in the $c=0$ limit, and where we have made no assumption about $\beta$. Like in IIA, \eqref{eq:bpsIIB3} indicates that we have a branching of possible solutions depending on how $\beta$ is tuned: this condition fixes $H_3$ such that its Bianchi identity is implied by the first of \eqref{eq:bpsIIB2}  unless $\cos\beta=0$. In this case $H_3$ is more weakly constrained by \eqref{eq:bpsIIB5}--\eqref{eq:bpsIIB7} and its Bianchi identity must be additionally imposed. We shall study these two cases in the next sections starting with $\cos\beta=0$. 

\subsection{Class III: Restricted almost K\"ahler foliation}\label{sec:IIBbetafixed}
In this section we consider the case where $\cos\beta=0$, we will actually fix $\beta=\frac{\pi}{2}$ without loss of generality.\\
~~\\
The analysis of this class runs pretty parallel to that of section \ref{sec:IIAbetafixed}. One can make progress solving \eqref{eq:bpsIIB2}--\eqref{eq:bpsIIB7} using the torsion classes of the SU(3)-structure \eqref{eq:torsions} and the identities they must obey \eqref{eq:id}. One finds in general that
\beq
A=A(y),~~~ \Phi=\Phi(y),
\eeq
and that the following torsion classes must vanish:
\beq\label{eq:classItorsionIIB}
V_2=\text{Im}E=T_1=T_2=W_1=\text{Re}W_2=W_3=W_4=W_5=0,
\eeq
which again makes M$_6$ a $y$-dependent almost K\"ahler manifold, of the restricted type we found in (\ref{eq:torsionsIIAcaseI}). The torsions $(\text{Re}E,S,\text{Im}W_2)$ remain non trivial, with the former fixed to
\beq\label{eq:classItorsionIIB0}
\text{Re}E= \frac{3m}{2}\frac{e^{A-\Phi}\sqrt{1- e^{-4A}h^2} \partial_y(e^{\Phi} h^2)}{h'(e^{4A}-2 h^2)},
\eeq
while the latter two appear in the definition of the NS flux:
\beq
H_3= e^{-2A}h\text{Re}S +\frac{ h h'}{m e^{3A}\sqrt{1- e^{-4A}h^2}}dy\wedge\text{Im}W_2+\frac{m e^{3A-\Phi}\sqrt{1- e^{-4A}h^2}}{2h h'(e^{4A}-2 h^2)}\partial_y(h^2 e^{\Phi})\text{Re}\Omega_3 .
\eeq
In addition to this, \eqref{eq:bpsIIB2}--\eqref{eq:bpsIIB7} imply one further condition on the functions of the ansatz, namely
\beq\label{eq:classIbpsIIB}
\partial_y\log\left(\frac{e^{2A-\Phi}}{\sqrt{1- e^{-4A}h^2}}\right)= 3 \frac{e^{-4A}hh'}{1- e^{-4A}h^2}.
\eeq  
One can  extract the magnetic components of the RR flux from \eqref{eq:reducedBPS7} which, given what has been derived thus far, may be expressed as
\begin{equation}
\begin{split}
	f_1&= -2 \frac{e^{-2A-\Phi}h}{\sqrt{1- e^{-4A}h^2}}d\log(e^{2A}h),\\[2mm]
	f_3&=\frac{e^{2A-\Phi}\sqrt{1- e^{-4A}h^2}}{h} H_3+\frac{m e^{A-\Phi}}{h}\text{Re}\Omega_3,~~~~f_5= \frac{e^{-2A-\Phi}h'}{\sqrt{1-e^{-4A}h^2}} dy\wedge J_2\wedge J_2,
\end{split}	
\end{equation}
where $f_7=0$ in general. Like the almost K\"ahler class in IIA, these fluxes imply both  \eqref{eq:reducedBPS6} and \eqref{eq:reducedBPS8}, so the supersymmetry equations are implied by just \eqref{eq:classItorsionIIB}, \eqref{eq:classItorsionIIB0} and \eqref{eq:classIbpsIIB}. What remains to ensure that one actually has a solution is to impose the Bianchi identity of the NS three-form. 

\subsection{Class IV:  A broad class}\label{sec:IIBbetagen}
In this section we assume that $\cos\beta \neq 0$, as we shall see this  leads to the least restrictive of our SU(3)-structure classes.\\
~~\\
For this class we find it helpful to decompose the exterior derivative as
\beq
d= d_6 + dy\wedge \partial_y.
\eeq
First off   \eqref{eq:bpsIIB4} implies that
\beq\label{eq:classIIIIBbps}
e^{2A-\Phi}\cos\beta= c_1, ~~~~ d\left(\tan\beta(1- e^{-4A}h^2)\right)+\frac{3}{2} h^2 \tan\beta d(e^{-4A})=0,
\eeq
where $c_1$ is a constant, and from \eqref{eq:bpsIIB3} we get that the NS flux is
\beq
H_3= d(\tan\beta J_2),
\eeq
solving its Bianchi identity by definition. Substituting  $(dJ_2,d\Omega_3)$ for their SU(3)-structure torsion classes in \eqref{eq:bpsIIB4}--\eqref{eq:bpsIIB7} it is possible to show that the only torsion classes that are necessarily trivial  are
\beq\label{eq:caseIIIIBtorsions}
\text{Im}E=\text{Re}W_1= \text{Re}W_2=0\,.
\eeq
Remarkably, this is a broader class than any of those in table \ref{table:one}; the requirements on M$_6$ are rather lax. We will find some simple explicit examples in later sections, but we expect that a lot more may emerge in the future with more effort.

The torsion forms $(T_2,S)$ remain completely arbitrary, while the rest are fixed in terms of these, $\alpha,\beta$ and the functions appearing in the metric ansatz as
\begin{align}\label{eq:caseIIIIBtorsions0}
\text{Re}E&= -\frac{m h^2(2-\cos(2\beta))\partial_y\log(e^{A}h)}{e^{3A}h'\sqrt{1- e^{-4A}h^2}},~~~\text{Re}V_2=-\frac{\cos\beta}{(1- e^{-4A}h^2)} d_6 (e^{-2A}h),
\\[2mm]
\text{Im}W_1&= -\frac{2m}{3}\frac{\cos\beta h}{e^{A}h'\sqrt{1- e^{-4A}h^2}}\partial_y\log(e^{A}h),~~~~\text{Im}W_2=-\frac{e^{2A}}{h\cos\beta}T_2,~~~~W_3=e^{-2A}h\cos\beta \text{Re}S,\nn\\[2mm]
W_4&=\frac{e^{-2A}h\sin^2\beta }{(1- e^{-4A}h^2)} d_6 (e^{-2A}h),~~~
\text{Re}W_5=-\frac{e^{-2A}h}{4(1- e^{-4A}h^2)}(e^{4A}h^{-2}+\cos(2\beta)-2)d_6 (e^{-2A}h),\nonumber
\end{align}
where the imaginary components of $(V_2,W_5)$ are implied since these are holomorphic one-forms. 
From \eqref{eq:reducedBPS7} we extract the magnetic fluxes
\begin{align}
f_1&= -\frac{2 c_1}{3 h\tan\beta}d(\sqrt{1-e^{-4A}h^2} \tan^2\beta),~~~~ f_5=-\frac{c_1\tan\beta h'}{e^{4A}\sqrt{1-e^{-4A}h^2}}dy\wedge J_2^2,\nn\\[2mm]
f_3&=\frac{c_1\sqrt{1-e^{-4A}h^2}}{h\sin\beta\cos\beta} H_3-\frac{c_1m}{e^{A}h \cos\beta}\text{Re}\Omega_3- \frac{2c_1 h'}{e^{4A}\sqrt{1- e^{-4A}h^2}}dy\wedge J_2, \label{eq:fluxIIBC2}\\[2mm]
f_7&=\frac{c_1 }{3 e^{4A}\sqrt{1-e^{-4A}h^2}}dy\wedge J_2^3. \nonumber
\end{align}
Some care should be taken with the expression for $f_3$ when $\sin\beta=0$, as while $H_3=0$ this is not true of  $(\sin\beta)^{-1} H_3$ (which is however finite). Again \eqref{eq:reducedBPS6} and \eqref{eq:reducedBPS8} are implied, so solving the supersymmetry constraints amounts to finding a manifold consistent with \eqref{eq:caseIIIIBtorsions}, \eqref{eq:caseIIIIBtorsions0} and a solution to the $\beta$ constraint in \eqref{eq:classIIIIBbps}; as the NS flux is already closed by definition there are no further conditions to be solved. We note that $\beta=0$ is again a potentially interesting limit, where this time the flux reduces to that of a D1--D5 system. We shall in fact recover the standard D1--D5 near horizon from this class in section \ref{sec:D1D5nearhoizon}, while we present a class generalising this and some new solutions in section \ref{ssub:S2-IIB}.\\
~~\\
This concludes our classification of type II solutions preserving ${\cal N}=(1,1)$ supersymmetry in terms of an SU(3)-structure. In the next section we shall construct some more refined classes within what we have presented, and find some old and new solutions.

\section{Solutions}\label{sec:solutions}
The aim of this section is to present some new solutions within the SU(3)-structure classes derived in sections \ref{sec:SU3IIA} and \ref{sec:SU3IIB}. We provide a summary of the physical solutions we find in section \ref{sub:summary}.\\
~~\\
Before presenting some new examples let us first perform a sanity check and recover something well known to preserve at least ${\cal N}=(1,1)$ supersymmetry in the next section.

\subsection{Recovering the D1--D5 near horizon}\label{sec:D1D5nearhoizon}
 An obvious solution that should lie within the SU(3)-structure classification of the preceding sections is the D1--D5 near horizon geometry, preserving ${\cal N}=(4,4)$. We know that this has a metric of the form
\beq\label{eq:D1D5met}
e^{2A}=L^2,\qquad ds^2(\text{M}_7)= e^{2C}ds^2(\text{S}^3)+ \lambda^2 ds^2(\text{CY}_2), 
\eeq
with $(e^{C},L,\lambda)$, and the dilaton set to a constant. Its only non trivial flux is the RR three-form, which has both an electric and magnetic component, so $f_3,f_7$ should be the only non trivial magnetic fluxes. As such this solution should lie within the class of section \ref{sec:IIBbetagen}, specialised to
\beq
\beta=0.
\eeq
As $dA=0$ for this solution, it is convenient to dispense with the local coordinate $y$: going back to (\ref{eq:localV}), we can write
\beq
h= L^2\cos\alpha,~~~~ V= \frac{L}{m} d\alpha.
\eeq
This makes it clear that we should take $\alpha$ to be a coordinate on M$_7$. The SU(3)-structure equations we need to solve are now
\begin{align}
dJ_2&=  \cos\alpha\text{Re}S+\frac{m}{\sin\alpha L}\text{Re}\Omega_3+d\alpha\wedge \left(\frac{2}{3}\cot\alpha J_2+\frac{L}{m} T_2\right),\nn\\[2mm]
d\text{Im}\Omega_3&=-\frac{2m}{3L\sin\alpha} J_2\wedge J_2-\frac{1}{\cos\alpha}J_2\wedge T_2+d\alpha\wedge \left(\frac{L}{m}\text{Im}S+\cot\alpha \text{Im}\Omega_3\right),\label{eq:D1D5ansatzSU3}\\[2mm]
d\text{Re}\Omega_3&=d\alpha\wedge \left(\frac{L}{m}\text{Re}S+\cot\alpha \text{Re}\Omega_3\right). \nonumber
\end{align}
Given that $V$ is aligned along $d\alpha$, it is natural to assume that is part of the three-sphere, so we decompose it as a foliation of S$^2$ over an interval:
\beq
ds^2(\text{S}^3)= d\alpha^2+ \sin^2\alpha ds^2(\text{S}^2),\qquad e^{2C}=\frac{L^2}{m^2}\,.
\eeq
We parameterise the two-sphere in terms of embedding coordinates $(y_1,y_2,y_3)$ such that $(y_a)^2=1$, so that we have the SU(2)  singlets\footnote{The D1--D5 near horizon supports small ${\cal N}=(4,4)$, as such it should support two sets of spinors charged under either a left or right SU(2) subgroup of the SO(4) global isometry of S$^3$. The specific SU(2) referenced here is the diagonal one formed of both these left and right SU(2)s}
\begin{align}
ds^2(\text{S}^2)=(dy_a)^2,~~~~\text{vol}(\text{S}^2)= \frac{1}{2}\epsilon_{abc}y_a dy_b\wedge dy_c
\end{align}
and the SU(2) triplets 
\beq
y_a,~~~~dy_a,~~~~ k_a=\epsilon_{abc} y_b dy_c,~~~~dk_a= 2 y_a \text{vol}(\text{S}^2).
\eeq
As there is a CY$_2$ factor in the metric, we have three closed two-forms, $(j_1,j_2,j_3)$ at our disposal obeying
\beq
j_a\wedge j_b=\frac{1}{2}\delta_{ab} \text{vol}(\text{CY}_2).
\eeq
A natural ansatz for the SU(3)-structure forms, giving rise to \eqref{eq:D1D5met} and obeying \eqref{eqSU3conds} is then
\beq\label{eq:D1D5Jomegaansatz}
J_2= \frac{L^2}{m^2}\sin^2\alpha \text{vol}(\text{S}^2)+ \lambda^2 y_a j_a,\qquad \Omega_3= \frac{L \lambda^2}{m}(dy_a- i k_a)\wedge j_a.
\eeq
Plugging this into \eqref{eq:D1D5ansatzSU3}, we find that it is solved by
\beq
S=0,\qquad T_2= \frac{4 L}{3m}\cos\alpha\sin\alpha\text{vol}(\text{S}^2)-\frac{2m \lambda^2}{3L}\cot\alpha y_a j_a.
\eeq
It is quick to show that $T_2$ is a primitive $(1,1)$-form, $T_2\wedge\Omega_3=T_2\wedge J_2\wedge J_2=0$, by introducing a complex vielbein on M$_6$; so the supersymmetry conditions are solved. Substituting the ansatz into \eqref{eq:fluxIIBC2} yields the fluxes
\beq\label{eq:D1D5flux}
f_3=\frac{2 c_1}{m^2} \sin^2\alpha d\alpha\wedge \text{vol}(\text{S}^2),~~~~f_7=-\frac{2 c_1 \lambda^4}{m^2}\sin^2\alpha d\alpha\wedge \text{vol}(\text{S}^2)\wedge \text{vol}(\text{CY}_2)
\eeq
with all else zero, so we have indeed reproduced the D1--D5 near horizon. The fact that this actually preserves ${\cal N}=(4,4)$ can be ascertained from the form of \eqref{eq:D1D5Jomegaansatz}. If we  rotate  $y_a$ in these expressions by a constant element of  SO(3), $(J_2,\Omega_3)$ change, but this new SU(3)-structure still solves \eqref{eq:D1D5ansatzSU3} (for a rotated $T_2$) and still gives rise to \eqref{eq:D1D5flux}. One can extract a further 3  independent SU(3)-structures in this fashion, for a total of 4 that the D1--D5 near horizon supports, so ${\cal N}=(1,1) \to (4,4)$.\\
~~\\
Having performed a non trivial check of the geometric conditions derived earlier, let us now construct some new solutions. We shall begin by assuming  M$_6$  is conformally $\text{CY}_3$.

\subsection{M$_6$ conformally Calabi--Yau}
The easiest way to find a solution within the new classes presented in  sections \ref{sec:SU3IIA}--\ref{sec:SU3IIB} is to impose that M$_6$ is (conformally) a Calabi--Yau three-fold (CY$_3$). This leads to four new local solutions, one for each class. In general we take the following ansatz for the metric 
\begin{align}\label{eq:ode-ccy-IIA-1}
ds^2&= e^{2A} ds^2(\text{AdS}_3)+V^2+ e^{2C} ds^2(\text{CY}_3),~~~~V=-\frac{1}{m\sqrt{1-e^{-4A}h (y)^2}} e^{-A}dy h'(y)
\end{align}
where we allow $e^{2C}$ to depend on $(y,\text{CY}_3)$ and decompose the SU(3)-structure forms as
\begin{equation}\label{eq:ode-ccy-IIA-2}
	 J_2= e^{2C}\hat J_2,\qquad \Omega_3= e^{3C}\hat \Omega_3,\qquad 
	d\hat J_2=d\hat\Omega_3=0.
\end{equation}
We begin by studying the possibilities of class I in section \ref{sec:IIAbetafixed}, then turn our attention to class III in section \ref{sec:IIBbetafixed}. CY$_3$ classes are also possible for class II and class IV, but they are special cases of more general warp nearly K\"ahler manifolds that we consider in section \ref{sec:NK}.

\subsubsection{CY$_3$ in IIA from class I}
Applying our conformal CY$_3$ ansatz to class I in section \ref{sec:IIAbetafixed}, we can find that all the torsion classes must vanish except for $\text{Re}E$  which enters the defining ODE for $C=C(y)$. Supersymmetry holds when the following ODEs are solved:
\beq
\partial_y\log\left(\frac{k}{h}\right)= \frac{3}{1- e^{-4A} h^2} \partial_y\log h,~~~~(1- e^{-4A}h^2)\partial_y C+\partial_y\log(e^{A}h)=0.
\eeq
For this class the Bianchi identity of the NS three-form is not implied, so we must impose it by hand. Away from the loci of possible NS sources this gives rise to the additional ODE
\beq
\frac{m e^{-A+3C}\partial_y\log(e^{A} h)}{\sqrt{1- e^{-4A}h^2}}= c \partial_y\log h,
\eeq
where $c$ is constant. A simple solution to this system is found by simply setting $ \partial_y C=0$, allowing us to integrate the ODEs with ease. By redefining
\beq
e^{2A}= \frac{h}{\cos\alpha}
\eeq 
where $\alpha=\alpha(y)$, we find the solution
\beq \label{eq:ode-ccy-IIA-sol}
h= L^2 \cos^{\frac{1}{3}}\alpha,~~~~k= L\lambda \frac{\cos^{\frac{4}{3}}\alpha}{\sin\alpha},~~~~ e^{2C}=1,~~~~ c=0,
\eeq
where $(L,\lambda)$ are integration constants. These imply that $H_3=0$ and at this point everything is a function of $\alpha(y)$, so we promote it to a coordinate. This is all that needs to be solved, the solutions takes the form
\begin{equation}
\begin{split}
	ds^2&= \frac{L^2}{\cos^{\frac{2}{3}}\alpha}\bigg[ds^2(\text{AdS}_3)+ \frac{1}{9m^2}d\alpha^2\bigg]+ ds^2(\text{CY}_3),\qquad e^{-\Phi}=\lambda \cos^{\frac{2}{3}}\alpha,\\[2mm]
	f_2&= \frac{m \lambda}{L} \hat{J}_2,\qquad f_4=-\frac{2\lambda}{3 }\cos^{\frac{2}{3}}\alpha d\alpha\wedge \text{Im}\hat\Omega_3,\qquad f_6=-\frac{m \lambda}{3!L}\hat{J}^3_2.	
\end{split}	
\end{equation}
The solution is bounded between  $0<\alpha\leq \frac{\pi}{2}$ and exhibits the following singular behavior at the extrema:
\beq\label{eq:cy2singularity}
ds^2 \sim L^2\bigg[\frac{1}{x^2}ds^2(\text{AdS}_3)+ \frac{x^2}{m^2}dx^2\bigg]+ ds^2(\text{CY}_3),
\eeq
where $|\alpha\pm\pi/2|= x^3$. This can be understood as the singularity associated to an intersection of several O-planes  extended in AdS$_3$ that wrap and are  smeared over various cycles in CY$_3$. \\
~\\
To see this, recall first that in general O-planes have a metric of the form 
\begin{equation}\label{eq:O}
	d s^2= H^{-1/2} d s^2_\parallel + H^{1/2} d s^2_\perp\,,
\end{equation}
where $H$ is a function of the transverse coordinates, harmonic with respect to $d s^2_\perp$. We now consider the cycle ${\mathcal C}$ in the CY$_6$ whose Poincar\'e dual is ${\hat J}_2$, and wrap a smeared O6 along it. This may consist of multiple components: for example when CY$_3=T^6$, ${\mathcal C}= \sum_{i=1}^3 {\mathcal C}_i$, where  ${\mathcal C}_1= T^2_{1234}$,  ${\mathcal C}_2= T^2_{1256}$, ${\mathcal C}_3= T^2_{3456}$, and in that case we are smearing three O6-planes wrapped along all three. On top of this set up one can place an O2 hole smeared over the whole of CY$_3$. For CY$_3=T^6$, the resulting metric takes the form
\begin{equation}\label{eq:4O}
\begin{split}
	ds^2=&\frac{1}{\sqrt{H_2 H_6^1H_6^2H_6^3}}ds^2(\text{AdS}_3)+\sqrt{H_2 H_6^1H_6^2H_6^3} dx^2\\
	&+ \sqrt{\frac{H^1_6 H_2}{H^2_6H^3_6}}ds^2({\cal C}_1)+ \sqrt{\frac{H^2_6 H_2}{H^1_6H^3_6}}ds^2({\cal C}_2)+ \sqrt{\frac{H^3_6 H_2}{H^1_6H^2_6}}ds^2({\cal C}_3),
\end{split}
\end{equation} 
the way the smearing has been performed means  the radial dependence of each warp factor  is the same:
\beq
H_2= H_6^1= H_6^2= H_6^3=  1- \frac{x}{x_0}.
\eeq     
Expanding about $x=x_0$ to leading order we reproduce the behavior of \eqref{eq:cy2singularity}. 

The metric singularity in (\ref{eq:cy2singularity}) is reproduced by any arrangement such that every M$_6$ direction is parallel to two smeared O-planes and transverse to two other O-planes; one should then look at the fluxes to decide what the correct O-plane system is. For supersymmetry one should also make sure that the number of directions that are transverse to one plane and parallel to another is always a multiple of four. Of course determining this precisely is not that important, since (partially) smeared O-planes such as these are not sensible in string theory.

\subsubsection{CY$_3$ in IIB from class III}
When we apply the conformal CY$_3$ ansatz to section \ref{sec:IIBbetafixed} we find again that $C=C(y)$, and that the supersymmetry and NS flux Bianchi identity require
\begin{align}
&\partial_y\log\left(\frac{e^{2A-\Phi}}{\sqrt{1- e^{-4A}h^2}}\right)= 3 \frac{e^{-4A}hh'}{1- e^{-4A}h^2},\qquad 2(e^{4A}-2h^2) \partial_y C+e^{-\Phi}\partial_y (e^{\Phi}h^2)=0,\nn\\[2mm]
&\frac{m e^{3A+3C-\Phi}\sqrt{1- e^{-4A}h^2}}{2h h'(e^{4A}-2 h^2)}\partial_y(h^2 e^{\Phi})=c,
\end{align}
for $c$ a constant. We again find a closed form solution when $\partial_y C=0$; it is rather similar to that of the previous section. Again substituting for $e^{2A}$ in favor of $\alpha$ through $e^{2A}= \frac{h}{\cos\alpha}$ we find that 
\beq
h=L^2\cos^{\frac{1}{3}} \alpha,~~~~ e^{-\Phi}=  \lambda \cos^{\frac{2}{3}}\alpha,~~~~e^{2C}=1,~~~~c=0,
\eeq
solves all the ODEs, setting $H_3=0$ in the process. The solution takes the form
\begin{equation}\label{eq:cy-IIB}
\begin{split}
	ds^2&= \frac{L^2}{\cos^{\frac{2}{3}}\alpha}\bigg[ds^2(\text{AdS}_3)+ \frac{1}{9m^2}d\alpha^2\bigg]+ ds^2(\text{CY}_3),\qquad e^{-\Phi}=\lambda \cos^{\frac{2}{3}}\alpha,\\[2mm]
	f_1&= \frac{2}{3}\lambda\cos^{\frac{2}{3}}\alpha d\alpha,\qquad f_3=-\frac{m \lambda}{L}\text{Re}\hat\Omega_3,\qquad f_5=\frac{\lambda}{3}\cos^{\frac{2}{3}}\alpha d\alpha\wedge \hat{J}^2_2.
\end{split}	
\end{equation}
We again find a solution bounded between $0\leq\alpha<\frac{\pi}{2}$ with the same singular behavior at the boundaries as in (\ref{eq:cy2singularity}). Clearly the interpretation must be a little different from (\ref{eq:4O}), as this time as O6- and O4-planes are not BPS in IIB. Inspecting the flux components suggests that the singularity should this time come from four intersecting O5-planes extendend in AdS$_3$ and wrapping four distinct three-cycles in CY$_3$ (for example, when CY$_3=T^6$, the four independent components of $\text{Im}\hat\Omega_3$). When such objects are smeared over the rest of CY$_3$ their warp factors are the same and reproduce the singularity.

As in IIA, the same metric singularity can be created by other arrangements of smeared O-planes, and one should examine the fluxes to determine the appropriate one; in future examples however we will refrain from doing so, because of the limited physical interest of smeared O-planes.\\

\subsection{M$_6$ conformally nearly K\"ahler }\label{sec:NK}
In this section we shall consider solutions that are foliations containing a nearly-K\"ahler 6 manifold NK$_6$:
\begin{align}
ds^2&= e^{2A} ds^2(\text{AdS}_3)+V^2+ e^{2C} ds^2(\hat{\text{M}}_6),~~~~V=-\frac{1}{m\sqrt{1-e^{-4A}h (y)^2}} e^{-A}dy h'(y)
\end{align}
where the SU(3)-structure forms are taken to be
\begin{equation}
\begin{split}
	 J_2&= e^{2C}\hat J_2,~~~~\Omega_3= e^{3C}\hat \Omega_3,\\[2mm]
	d\hat J_2&= 3n \text{Im}\hat\Omega_3,~~~~d\hat\Omega_3=-4n \hat J_2\wedge \hat J_2,
\end{split}	
\end{equation}
where $n$ is a constant. This ansatz is appropriate for classes II and IV in sections \ref{sec:IIAbetagen} and \ref{sec:IIBbetagen}. For a NK$_6$ manifold we can  actually fix $n= 1$ without loss of generality, however when $n=0$ we  have conditions for CY$_3$ solutions not considered in the previous section, so we shall keep $n$ arbitrary for now.

Only a few NK$_6$ manifolds are known so far: S$^6$, S$^3\times \mathrm{S}^3$, $\mathbb{CP}^3$, and the so-called flag manifold $\mathbb{F}(1,2;3)$. The latter two are homogeneous, while the first two admit both homogeneous and cohomogeneity-one nearly-K\"ahler structures \cite{foscolo-haskins}.

\subsubsection{IIA solutions}
Applying the ansatz of the previous section to the class of section \ref{sec:IIAbetagen} we find that solutions in IIA are governed by the following ODEs:
\begin{equation}
\begin{split}
	&\partial_y\left(\frac{e^{2A+6C}h^2}{\cos^2\beta}\right)=\frac{2n e^{A-C}}{m \cos\beta}\partial_y\left(\frac{e^{2A+6C}h^2\sqrt{1- e^{-4A}h^2}}{\cos^2\beta}\right),\\[2mm]
	&\partial_y\left(\frac{e^{2A+6Ch^2}\sqrt{1- e^{-4A}h^2}}{\cos^2\beta}\right)=-3 \frac{e^{-2A+6C}h^3h'}{\cos^2\beta\sqrt{1- e^{-4A}h^2}},~~~~\partial_y\left(\frac{e^{6C}\sin\beta}{\cos^3\beta}\right)=0,\label{eq:NK6IIAbps}
\end{split}
\end{equation}
where $C=C(y)$. These imply supersymmetry and the Bianchi identities of the fluxes.\\
~~\\
\textbf{CY$_3$ limit}\\
~\\
First off, it is a simple matter to establish that when $n=0$, the CY$_3$ limit, the ODEs truncate to
\beq
\partial_y(e^{A}h)=0,~~~\partial_yC=\partial_y\beta=0,
\eeq
which can be solved as
\beq
h=L^2\cos^{\frac{1}{3}} \alpha,~~~e^{2C}=1,~~~\beta=\text{constant}
\eeq
where we have again made use of $e^{2A}\cos\alpha= h$.
 This gives rise to the following analytic CY$_3$ solution
\begin{align}
ds^2&= \frac{L^2}{\cos^{\frac{2}{3}}\alpha}\bigg[ds^2(\text{AdS}_3)+ \frac{1}{9m^2}d\alpha^2\bigg]+ ds^2(\text{CY}_3),\qquad e^{-\Phi}=\frac{c_1 L\cos^{\frac{2}{3}}\alpha}{\cos\beta},\nn\\[2mm]
f_0&= c_1 m,\qquad f_2= c_1m \tan\beta \hat{J}_2,\qquad f_6=-\frac{c_1 m}{3!}\tan\beta \hat J_2^3,\\[2mm]
f_4&= -\frac{c_1m}{2}\hat J_2\wedge \hat J_2-\frac{2 c_1 L\cos^{\frac{2}{3}}\alpha}{3 \cos\beta}d\alpha\wedge \text{Im}\hat\Omega_3,
\nonumber
\end{align}
We see that the solution is bounded between singularities of the type discussed below (\ref{eq:4O}); this time it can be explained as the intersection of a D8/O8 system and three O4 planes that each wrap a distinct two-cycle in CY$_3$ and are smeared over the rest of it. \\
~\\
\textbf{NK$_6$ solutions}\\
~\\
When $n\neq 0$, M$_6$ is a nearly-K\"ahler manifold. We take $n=1$, without loss of generality.
In this case we have not been able to find any analytic solutions to \eqref{eq:NK6IIAbps}. One can make progress by studying it in a power series expansion. We have found it easier to study the system after fixing the $y$ coordinate reparameterization freedom by demanding $h'= \sqrt{1-h^2 e^{-4A}}$.\\
~\\
For the solution to be compact, $y$ needs to belong to an interval $[y_-,y_+]$, and one or more of the other directions should shrink at its endpoints $y_\pm$. If we take $e^{C(y_+)}=0$, M$_6$ shrinks; this is a regular point when M$_6=$S$^6$, and is a conical G$_2$ singularity otherwise, which is believed to be allowed in string theory. By assuming a power series with integer coefficients for all the relevant functions, we find (for $y_+=0$) 
\begin{equation}\label{eq:cg2}
	e^A = a_0 - \frac3{49a_0^3}y^2 + O(y)^4 \, ,\qquad
	e^C = -\frac1{m a_0}y -\frac5{98 m a_0^5} y^3 + O(y)^5 
	\, ,\qquad
	h = a_0^2 - \frac3{14 a_0^2}y^2 + O(y)^4\,.
\end{equation}
A standard procedure is to evaluate this at a small value of $y$, and evolve it numerically (towards negative $y$). This ends with a singularity, which upon further numerical inspection is revealed to be
\begin{equation}\label{eq:O8}
	d s^2\sim (y-y_-)^{-1/2} ds^2(\text{AdS}_3) + (y-y_-)^{-1/2} ds^2(\text{M}_6) + (y-y_-)^{-1/2} dy^2 \quad \text{[O8]}\,.
\end{equation}
Comparing with (\ref{eq:O}), we see that this can be straightforwardly identified as an O8-plane, since in that case $H$ is linear.\\ 
~\\
\begin{figure}[ht]
	\centering
		\includegraphics[width=10cm]{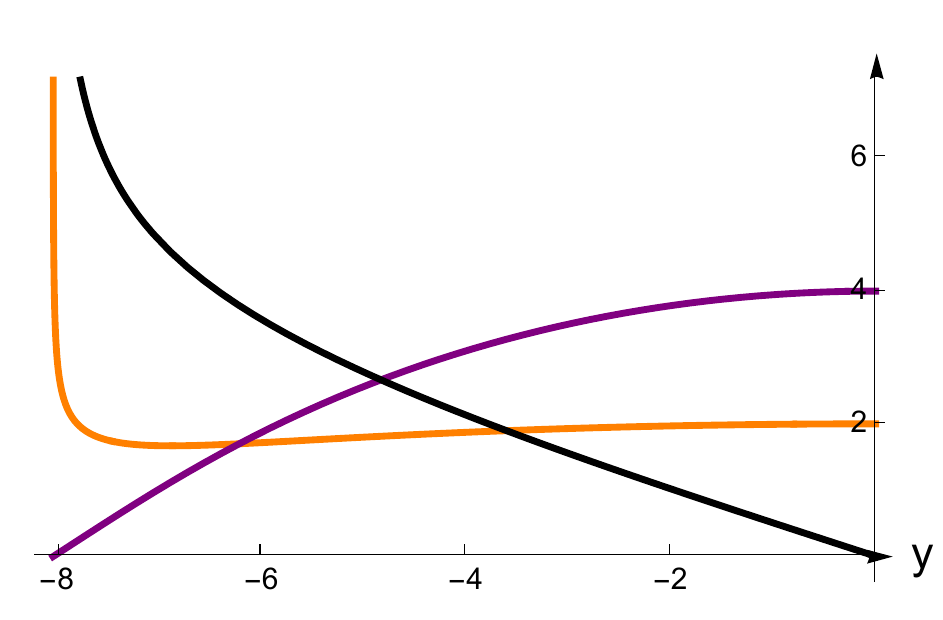}
	\caption{\small A numerical solution with M$_6$ nearly-K\"ahler, interpolating between an O8-plane on the left, and a regular (or conical G$_2$ singularity) on the right. The functions are $e^A$ (orange), $e^C$ (black), $h$ (purple). The parameters in (\ref{eq:cg2}) in this case are $a_0=2$, $m=1$.}
	\label{fig:nkIIA-O8-reg}
\end{figure}
We have also studied the system (\ref{eq:NK6IIAbps}) numerically, by evolving from random initial conditions at values $y=y_0$ towards both smaller and larger $y$. The evolution continues on both sides until it stops at two types of singularities, which again can be further investigated by zooming around them. One type of solutions we obtained this way has an O8-plane (\ref{eq:O8}) at the endpoint $y=y_-$, and 
\begin{equation}\label{eq:smO2}
	d s^2\sim (y-y_+)^{-1/2} ds^2(\text{AdS}_3) + (y-y_+)^{1/2} ds^2(\text{M}_6) + (y-y_+)^{1/2} dy^2 \quad \text{[O2]}\,
\end{equation} 
at $y=y_+$. Recalling again (\ref{eq:O}), we see that this has the structure of an O2-plane: $y$ is interpreted as the radial direction on the cone $C(\text{M}_6)$ over M$_6$, and the harmonic function $H\sim 1 - (r_0/r)^5$, expanded around $r=r_0$, produces a linear $H\sim y$. This is is an ordinary O2-plane when M$_6= $S$^6$, and otherwise represents an O2 at the tip of a conical G$_2$ singularity.\\ 

~\\
A final, less meaningful type of numerical solutions has again an O8-plane at $y=y_-$, and 
\begin{equation}\label{eq:3O4}
	ds^2 \sim (y-y_0)^{-3/2} ds^2(\text{AdS}_3) + (y-y_0)^{1/2} ds^2(\text{M}_6) + (y-y_0)^{3/2} dy^2\quad \text{[3 smeared O4-planes]}\,.
\end{equation}
This can be interpreted along similar lines to (\ref{eq:4O}), as the result of three smeared O4-planes. Again all are parallel to AdS$_3$ and perpendicular to $dy$; one extended along directions $x^1$, $x^2$; one along $x^3$, $x^4$; one along $x^5$, $x^6$, in local coordinates on M$_6$.

\subsubsection{IIB solutions}
Applying the  NK$_6$ ansatz to the class of section \ref{sec:IIBbetagen} we find that solutions in IIB are in one to one correspondence with the following ODEs:
\begin{equation}\label{eq:NK6IIBbps}
\begin{split}
	&\partial_y\left(\frac{e^{6C}}{\cos^2\beta}\right)=\frac{2n e^{-A+C}}{m \cos\beta}\partial_y\left(\frac{e^{6C}\sqrt{1- e^{-4A}h^2}}{\cos^2\beta}\right),\\[2mm]
	&\partial_y\left(\frac{e^{6C}\sqrt{1- e^{-4A}h^2}}{\cos^2\beta}\right)=-3 \frac{e^{6C-4A}h h'}{\cos^2\beta\sqrt{1- e^{-4A}h^2}},~~~~\partial_y\left(\frac{e^{6C}\sin\beta}{\cos^3\beta}\right)=0,
\end{split}	
\end{equation}
where again $C=C(y)$.\\
~~\\
\textbf{CY$_3$ limit}\\
~\\
As in IIA, when we fix $n= 0$ we get CY$_3$ solutions, as before they are unwarped  and an analogous calculations yields
\begin{align}
ds^2&= \frac{L^2}{\cos^{\frac{2}{3}}\alpha}\bigg[ds^2(\text{AdS}_3)+ \frac{1}{9m^2}d\alpha^2\bigg]+ ds^2(\text{CY}_3),~~~~ e^{-\Phi}=\frac{c_1 \cos^{\frac{2}{3}}\alpha}{L^2\cos\beta},\nn\\[2mm]
f_1&= -\frac{2c_1 \tan\beta}{3L^2}\cos^{\frac{2}{3}}\alpha d\alpha,~~~~f_5= \frac{c_1}{3 L^2} \tan\beta\cos^{\frac{2}{3}}\alpha d\alpha\wedge \hat{J}^2_2,~~~~f_7=-\frac{c_1}{9 L^2}\cos^{\frac{2}{3}}\alpha d\alpha\wedge\hat J_2^3,\nn\\[2mm]
f_3&= \frac{c_1m}{L^3\cos\beta}\text{Re}\hat\Omega_3+\frac{2c_1}{3L^2}\cos^{\frac{2}{3}}\alpha d\alpha\wedge \hat J_2.
\end{align}
This time the singularity can be interpreted as that coming from four O5-planes that wrap four distinct 3-cycles in CY$_3$ and are smeared over the rest of it.\\ 
~\\
\textbf{NK$_6$ solutions}\\
~\\
In this case we have not found any analytic solutions. While it is possible to find a local regular (or conical G$_2$) solution similar to (\ref{eq:cg2}), its numerical evolution ends at points without a clear physical interpretation. The same issue presents itself when evolving from a random point internal to the $y$ interval.

\subsection{Foliations over Tw(M$_4$)}
A natural generalisation of the D1--D5 near horizon SU(3)-structure \eqref{eq:D1D5Jomegaansatz} is to gauge the SU(2) of the two-sphere, such that our ansatz for the internal seven-manifold becomes 
\begin{equation}\label{eq:tw-met}
\begin{split}
	ds^2(\text{M}_7)&=V^2+ds^2(\text{M}_6),~~~~ V=-\frac{1}{m\sqrt{1-e^{-4A}h (y)^2}} e^{-A}dy h'(y)\\[2mm]
	ds^2(\text{M}_6)&= e^{2B}Dy_aDy_a+ e^{2C} ds^2(\text{M}_4),~~~~ Dy_a =dy_a + \epsilon_{abc}y_bA_c	
\end{split}	
\end{equation}
where we take $(e^A,e^{B},e^C)$ to be functions of $y$ only. We take $A_i$ to be the connection on the bundle of anti-self-dual forms on M$_4$; this makes M$_6$ the twistor bundle Tw(M$_4$). A basis of two-forms $j_i$ on M$_4$ then satisfies
\beq
dj_a= -\epsilon_{abc} A_b\wedge j_c.
\eeq
We can introduce an SU(3)-structure as
\beq
J_2= \frac{1}{2}e^{2B}\epsilon_{abc}y_a Dy_b \wedge Dy_c + e^{2C} y_a j_a,~~~~\Omega_3= e^{B+2C}(dy_a- i \epsilon_{abc}y_b Dy_c)\wedge j_a.
\eeq
We shall also assume that M$_4$ is Einstein and self-dual.\footnote{There are only two smooth such manifolds, $S^4$ and $\mathbb{CP}^2$ \cite{hitchin-qK}, but many more with orbifold singularities \cite{boyer-galicki-book}.} This implies that the field strength $F_a= dA_a+ \frac{1}{2} \epsilon_{abc}A_b\wedge A_c$ is proportional to $j_a$:\footnote{For more details on this set-up, see for example \cite[Sec.~9.2]{Gauntlett:2006ux}, or \cite{Tomasiello:2007eq} in a different language for AdS$_4$ solutions. M$_6$ is half-flat more generally when M$_4$ is only self-dual and not Einstein, but in this case we have not been able to solve the remaining conditions on M$_7$. It would be very interesting to find solutions with general M$_4$; we thank E.~Witten for related discussions.}
\beq \label{eq:fa=ja}
F_a = b j_a,~~~~ db=0,
\eeq
so that we shall also recover conformal CY$_2$ solutions if we fix $b=0$. More generically $b$ is proportional to the constant curvature on M$_4$, $b>0$ is positive curvature, $b<0$ negative. There are two cases where it makes sense to try this ansatz, namely the $\beta=\beta(y)$, $d_6\alpha=0$ limit of cases II and IV.

\subsubsection{IIA solutions}\label{ssub:S2-IIA}
\begin{figure}[ht]
	\centering
		\includegraphics[width=10cm]{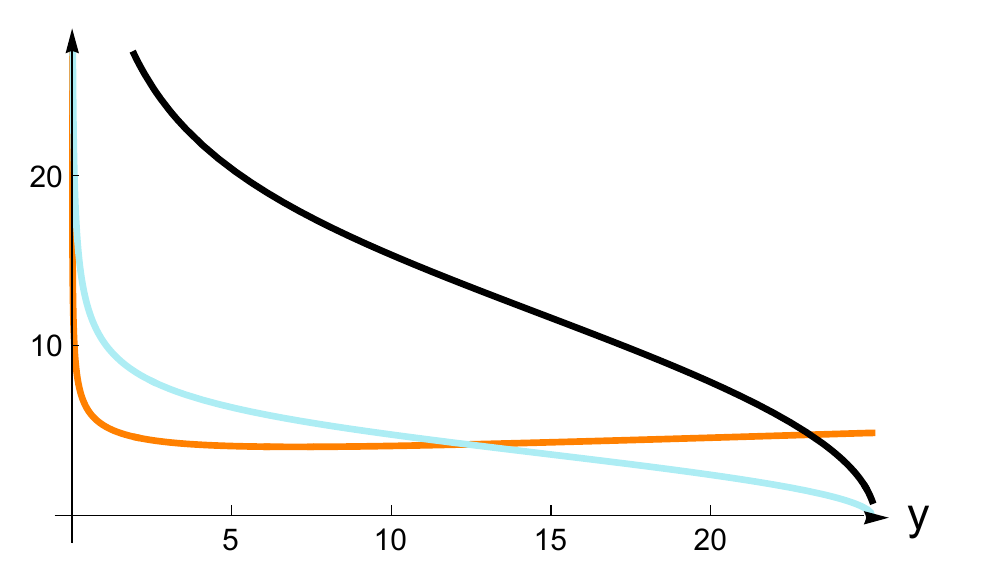}
	\caption{\small A numerical solution with M$_6=\mathrm{Tw}(\mathrm{M}_4)$, interpolating between an O8-plane on the left, and a regular point on the right. The functions are $e^A$ (orange), $e^B$ (aqua), $e^C$ (black). The parameters in (\ref{eq:reg-S2-IIA}) are $a_0=5$, $c_0=5$, $b=1$, $m=1$.}
	\label{fig:S2IIA-O8-reg}
\end{figure}

Plugging this into the torsion classes of section \ref{sec:IIAbetagen} we find
\beq\label{eq:twisterTIIA}
S=0,~~~~T_2= \frac{2}{3} \cos\beta\frac{e^{2A+B}}{h}\bigg(e^{-2B}-2b e^{-2C}\bigg)\bigg(\frac{1}{2}e^{2B}\epsilon_{abc}y_aDy_b\wedge Dy_c- e^{2C} y_a j_a\bigg)
\eeq
where $T_2$ is indeed a primitive (1,1)-form. We additionally get the following ODEs
\begin{equation}\label{eq:S2-IIA-odes}
\begin{split}
\partial_y\left(\frac{e^{2B+2C}}{\cos^2\beta}\right)&=-\frac{2}{m h}\frac{e^{A+B+2C}h'}{\cos\beta\sqrt{1- e^{-4A} h^2}},~~~~\partial_y \left(\frac{e^{4C}}{\cos^2\beta}\right)=2 b \partial_y\left(\frac{e^{2B+2C}}{\cos^2\beta}\right),\\[2mm]
\partial_y (e^{A}h)&=\frac{e^{2A}h'(e^{2C}+b e^{2B})\sqrt{1- e^{-4A}h^2}}{m e^{B+2C}\cos\beta },~~~~~\partial_y\left(\frac{e^{2B+4C}\sin\beta}{\cos^3\beta}\right)=0,
\end{split}	
\end{equation}
which imply supersymmetry and the EOM. When $\beta=0$, the fluxes are those of a D4--D8 system with Romans mass $F_0= c_1 m$.\\
~\\
We did not find any analytic solutions to (\ref{eq:S2-IIA-odes}). As in \ref{sec:NK}, one can find local solutions with a power series expansion. In this case it is slightly easier to take the radial coordinate to be $y=h$ itself.\\
~~\\
A regular endpoint $y_+$ of the $y$ interval can now be obtained by making the S$^2$ shrink, which is achieved by imposing $e^B(y_+)=0$. We find $y_+=a_0^2$ and
\begin{align}\label{eq:reg-S2-IIA}
	\nonumber e^A &= a_0 + O(a_0^2-y)^2 \, ,\qquad
	e^B = \sqrt{2}m (a_0^2-y)^{1/2} + \frac{7m^2c_0^2-8b a_0^2}{10\sqrt{2} m^3 a_0^2 c_0^2}(a_0^2-y)^{3/2}+ O(a_0^2-y))^{5/2}
	\, ,\\
	e^C &= c_0 + \frac{b}{c_0 m^2} (a_0^2-y) + O(a_0^2-y)^{2}\,.
\end{align}
A numerical evolution starting from this local solution ends with a singularity at $y=y_-$. For $b>0$ this is again the O8-plane singularity (\ref{eq:O8}), so this gives another class of solutions where an O8 is the only O-plane; we show an example in Fig.~\ref{fig:S2IIA-O8-reg}. For M$_4= $S$^4$ and $\mathbb{CP}^2$, the $\mathrm{Tw}(\text{M}_4)$ is $\mathbb{CP}^3$ and a flag manifold respectively; both admit nearly-K\"ahler metrics. However, even these cases are distinct from the solutions in section \ref{sec:NK}, as we see by comparing the local solution (\ref{eq:reg-S2-IIA}) with (\ref{eq:cg2}). In particular, away from the O8 the present ones are fully regular for any M$_4$, while those of section \ref{sec:NK} are regular only when M$_6=$S$^6$, and have a conical G$_2$ singularity otherwise. \\
~\\
For $b<0$, a numerical evolution from (\ref{eq:reg-S2-IIA}) ends with the singularity at $y=y_-$:
\begin{equation}\label{eq:smO4}
	d s^2\sim (y-y_-)^{-1/2}\left( ds^2(\text{AdS}_3) + D y^a Dy^a\right) + (y-y_-)^{1/2} (ds^2(\text{M}_4) + dy^2) \quad \text{[(smeared) O4]}\,.
\end{equation}
Comparing with (\ref{eq:O}), this would appear to have the structure of an O4, with $H\sim y$. At this point one might try to interpret $y$ as the radial direction on the cone $C(\text{M}_4)$ over M$_4$; a harmonic function $h\sim 1 - (r_0/r)^3$, expanded around $r=r_0$, would produce a linear $h\sim y$. This cone $C(\text{M}_4)$ would be smooth for M$_4=$S$^4$, but unfortunately this is excluded by the assumption $b<0$. For other choices of four-manifolds,  $C(\text{M}_4)$ is not a physically sensible singularity. Another possible interpretation is that $y$ is not radial, but one of the transverse coordinates; (\ref{eq:smO2}) is then an O4 partially smeared along M$_4$. This is unfortunately not really sensible in string theory, although one may hope that it points to the existence of a solution where the O4 is fully localized.\footnote{See \cite[Sec.~4.1.1]{Passias:2018zlm} for a more extended discussion of the difference between genuine and smaread O-plane singularities.}\\
~~\\
Globally, we have been able to obtain numerical solutions where $y\in [y_-,y_+]$, that behave as O2-planes (\ref{eq:smO2}) on both sides, and others that behave as an O2 on one side and an O8 on the other.

\subsubsection{IIB solutions}\label{ssub:S2-IIB}
Plugging the twistor ansatz (\ref{eq:tw-met})--(\ref{eq:fa=ja}) into the torsion classes of section \ref{sec:IIBbetagen} we find
\beq
S=0,~~~~T_2= \frac{2}{3}h e^{-2A+B}\cos\beta\bigg(e^{-2B}-2b e^{-2C}\bigg)\bigg(\frac{1}{2}e^{2B}\epsilon_{abc}y_aDy_b\wedge Dy_c- e^{2C} y_a j_a\bigg),
\eeq
which differs from \eqref{eq:twisterTIIA} only by a pre-factor.  $T_2$ is again a primitive (1,1)-form, we also get the following ODEs
\begin{equation}\label{eq:ode-S2-IIB}
\begin{split}
	&\partial_y\left(\frac{e^{2B+2C}}{\cos^2\beta}\right)=-\frac{2h}{m }\frac{e^{B+2C}h'}{e^{3A}\cos\beta\sqrt{1-e^{-4A}h^2}},~~~~\partial_y\left(\frac{e^{4C}}{\cos^2\beta}\right)=2b\partial_y\left(\frac{e^{2B+2C}}{\cos^2\beta}\right),\\[2mm]
	&\partial_y(e^{A}h)=\frac{e^{2A}h'(b e^{2B}+e^{2C})\sqrt{1- e^{-4A}h^2}}{m e^{B+2C}\cos\beta},~~~~\partial_y\left(\frac{e^{2B+4C}\sin\beta}{\cos^3\beta}\right)=0	
\end{split}
\end{equation}
which imply supersymmetry and all the EOM. Note that fixing $b=0$ (so that the fibration becomes topologically trivial) gives a generalisation of the D1--D5 near horizon, additionally fixing $e^A$=constant. $\beta=0$ reduces to it.\\
~\\
We applied to (\ref{eq:ode-S2-IIB}) the same procedure we saw in IIA. We did manage to find a regular local solution similar to (\ref{eq:reg-S2-IIA}), but numerical evolution from it now ends with a singularity of the type (\ref{eq:cy2singularity}), with the CY$_6$ replaced by M$_6$. Its possible interpretation in terms of smeared O-planes was discussed there and below (\ref{eq:cy-IIB}).
~\\
Finally, numerical evolution from a random internal point results in solutions which have two solutions of the type (\ref{eq:cy2singularity}) at both endpoints $y_\pm$.

\subsection{S$^3$ fibered over $\Sigma_3$}
In this section we shall consider solutions where M$_6$ is S$^3$ fibered over a 3-manifold of constant curvature.  We shall take the metric to be
\begin{equation}
\begin{split}
	ds^2(\text{M}_7)&=V^2+ds^2(\text{M}_6),~~~~ V=-\frac{1}{m\sqrt{1-e^{-4A}h (y)^2}} e^{-A}dy h'(y)\\[2mm]
	ds^2(\text{M}_6)&= e^{2B}(\mu^i)^2+ e^{2C} ds^2(\Sigma_3),~~~~ \mu_i =\omega_i + A_i,~~~~d\omega_i=\frac{1}{2}\epsilon_{ijk}\omega^j\wedge \omega^k,
\end{split}	
\end{equation}
where $(e^{2B},e^{2C})$ depend on $y$ and the connection $A_i$ is defined on $\Sigma_3$, which we take to have constant curvature. We will impose that the SU(2) field strength of the connection is
\beq
F_i= -\frac{R_0}{12}\epsilon_{ijk}e^j\wedge e^k
\eeq
where $R_0$ is the Ricci scalar on $\Sigma_3$. We will take the SU(3)-structure forms to be
\begin{align}
J_2&= e^{B+C}\mu^i\wedge e^i,\nn\\[2mm]
\text{Re}\Omega_3&=e^{3C}e^1\wedge e^2\wedge e^3-\frac{1}{2}e^{2B+C}\epsilon_{ijk}\mu^i\wedge \mu^j\wedge e^k,\\[2mm]
\text{Im}\Omega_3&=e^{3B}\mu_1\wedge \mu_2\wedge \mu_3-\frac{1}{2}e^{B+2C}\epsilon_{ijk}\mu^i\wedge e^j\wedge e^k.\nonumber
\end{align}

\subsubsection{IIA solutions}\label{ssub:S3-IIA}
\begin{figure}[ht]
	\centering
		\includegraphics[width=10cm]{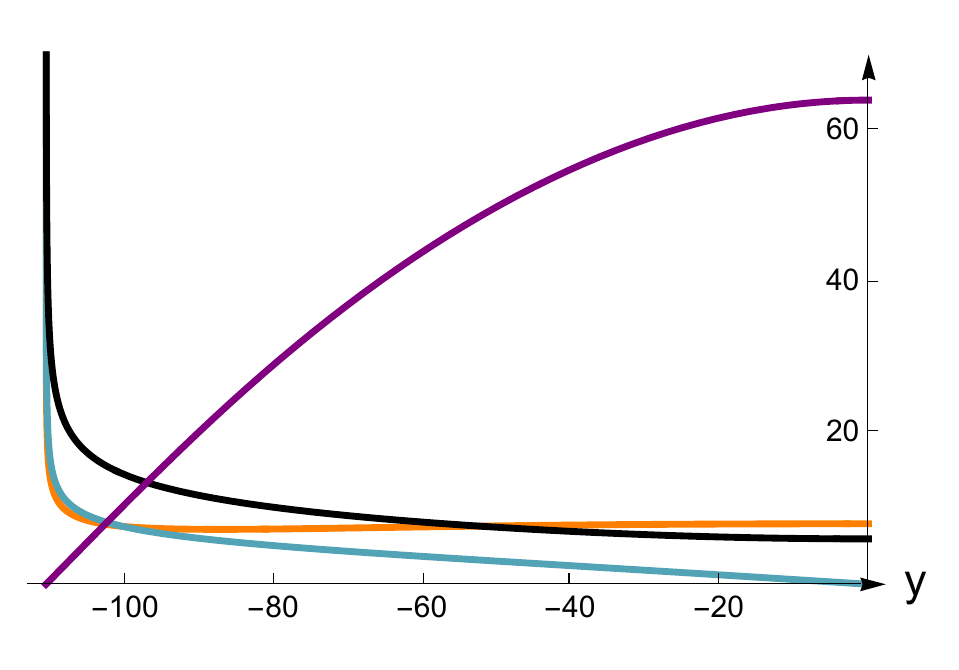}
	\caption{\small A numerical solution with M$_6$ an $\mathrm{S}^3$-fibration over a three-manifold $\Sigma_3$, interpolating between an O8-plane on the left, and a regular point on the right. The functions are $e^A$ (orange), $e^B$ (aqua), $e^C$ (black), $h$ (purple). The parameters in (\ref{eq:reg-S3-IIA}) in this case are $a_0=8$, $c_0=4$, $R_0=6$, $m=1$.}
	\label{fig:S3IIA-O8-reg}
\end{figure}

Plugging the ansatz into for the SU(3)-structure of the previous section into the classification of section \ref{sec:IIAbetagen} we find that the non trivial torsion classes are fixed as
\begin{align}
T_2&=0,~~~~~S=\frac{e^{-2A-2B-C}h(R_0e^{2C}-2e^{2B})}{8\cos\beta}\bigg[-3e^{3C}e^1\wedge e^2\wedge e^3-\frac{1}{2}e^{2B+C}\epsilon_{ijk}\mu^i\wedge \mu^j\wedge e^k\nn\\[2mm]
&+i\big(3e^{3B}\mu_1\wedge \mu_2\wedge \mu_3+\frac{1}{2}e^{B+2C}\epsilon_{ijk}\mu^i\wedge e^j\wedge e^k\big)\bigg]
\end{align}
where $S$ is indeed a primitive (2,1)-form. In addition to this we get the following ODEs for the functions of the ansatz:
\begin{align}
\partial_y\left(\frac{e^{A+3C}h}{\cos\beta}\right)&=\frac{R_0}{2}\partial_y\left(\frac{e^{A+2B+C}h}{\cos\beta}\right),~~~~
\partial_y\left(\frac{e^{A+2B+C}h}{\cos\beta}\right)=\frac{1}{m}\frac{e^{-2A+B+C}h^2h'}{\cos^2\beta \sqrt{1- e^{-4A}h^2}} ,\nn\\[2mm]
\label{eq:ode-S3-IIA}
\partial_y\left(\frac{e^{2B+2C}}{\cos^2\beta}\right)&= \frac{1}{6m}\frac{e^{A+B}(e^{2B}R_0+6 e^{2C})h'}{\cos\beta h \sqrt{1-e^{-4A}h^2}},~~~~\partial_y\left(\frac{e^{3B+3C}\sin\beta}{ \cos^3\beta}\right)=0.
\end{align}
These imply that
\beq
e^{3B+3C}\sin\beta= b_1  \cos^3\beta,\qquad e^{A+3C}h-\frac{R_0}{2}e^{A+2B+C}h=b_2\cos\beta
\eeq
where $b_{1,2}$ are arbitrary integration constants.\\
~~\\
We find the following solution to these conditions:\footnote{Positivity of the metric fixes $R_0<0$, but it can then be set to $-6$ without loss of generality.}
\begin{align}\label{eq:S3-analytic-sol}
\beta&=0,~~~~ R_0=-6,\\[2mm]
e^{2B}&=\frac{4}{9m^2}e^{2A}(1- e^{-4A}h^2),~~~~
e^{2C}=\frac{4}{3 m^2}e^{2A},~~~~
he^{4A}=\frac{1}{2}(C+h^3),~~~~dC=0.\nn
\end{align}
We can make more sense of the solution by fixing
\beq
C= 2^{\frac{3}{2}}L^6,\qquad  h= \sqrt{2}L^2 \cos^{\frac{2}{3}} y,\qquad m c_1=F_0.
\eeq
The metric and dilaton then become
\begin{align}
ds^2&=\frac{L^2}{\cos^{\frac{1}{3}} y}\bigg[\sqrt{\Delta}\bigg(ds^2(\text{AdS}_3)+\frac{4}{3m^2}ds^2(\mathbb{H}_3)\bigg)+\frac{16}{9m^2 \sqrt{\Delta}}\bigg(\frac{1}{2}\Delta dy^2+\frac{1}{4}\sin^2y(\mu^i)^2\bigg)\bigg],\nn
\label{eq:S3-analytic}\\[2mm]
e^{-\Phi}&=\frac{\sqrt{2}F_0 L\cos^{\frac{5}{6}}y}{m \Delta^{\frac{1}{4}}},~~~~\Delta= 1+\cos^2y.
\end{align}
If we ignore the overall warping, the $(y,\mu_i)$ directions are topologically a 4-sphere with $0<y<\pi$, however this gets restricted to the hemisphere $0<y<\frac{\pi}{2}$ in the full space. One can easily see that $y=0$ is regular zero while $y=\frac{\pi}{2}$ is a singular locus; this is in fact what one expects for  D8 branes coincident to an O8 plane extended in all by the $y$ direction. This suggests that we have derived a solution with a system of D4--D8 branes compactified on $\mathbb{H}_3$ (mod some discrete subgroup). In other words, this solution should be the uplift of the AdS$_3\times \mathbb{H}_3$ vacua of Romans F(4) found in \cite{Nunez:2001pt}; an uplift of this solution to IIB was also recently performed in \cite{Legramandi:2021aqv}. The non trivial fluxes are the Romans mass $F_0$ and the 4-form which is purely magnetic, given by
\beq
F_4= dC_3,~~~ C_3= \frac{8 L^4 F_0}{27m^3}\cos^{\frac{4}{3}} y\bigg(4\left(\frac{1}{8}-\frac{1}{\Delta}\right)\mu^1\wedge\mu^2\wedge \mu^3+\frac{3}{4}\epsilon_{ijk}e^i\wedge e^j\wedge \mu^k \bigg).
\eeq
~\\
The analytic solution (\ref{eq:S3-analytic}) is not the most general solution to the system (\ref{eq:ode-S3-IIA}). As we did in previous sections, we can look for more general solutions by imposing regularity at one of the endpoints, by demanding that the $\mathrm{S}^3$ shrink there (just like $y=0$ in (\ref{eq:S3-analytic})). This leads to a more general class, 
which we give in the gauge where $h'= \sqrt{1-h^2 e^{-4A}}$: 
\begin{equation}\label{eq:reg-S3-IIA} 
\begin{split}
	e^A&= a_0-\frac3{64 a_0^3} y^2 + O(y)^4 \, ,\qquad
	e^B = \frac1{2a_0 m} y + \frac{9 m^2 c_0^2 - 8 a_0^2 R_0}{1152 a_0^5 c_0^2 m^3}y^3+ O(y)^5\,\\
	e^C &= c_0 +\frac{8 a_0^2 R_0 + 27 c_0^2 m^2}{192 a_0^4 c_0 m^2} y^2 + O(y)^4\, ,\qquad h= a_0^2 -\frac3{8a_0^2}y^2 + O(y)^4
	 \,.
\end{split}	
\end{equation}
For $c_0 = \frac{2a_0}{\sqrt{3} m}$, this is an expansion of (\ref{eq:S3-analytic-sol}). A numerical evolution of this perturbative solution often results in an O8-plane singularity (\ref{eq:O8}), just as in (\ref{eq:S3-analytic}). So that solution has a two-parameter numerical generalization, also for $R_0>0$ ($\Sigma_3 = \mathrm{S}^3$). We show one numerical example in Fig.~\ref{fig:S3IIA-O8-reg}.\\
~~\\
For $R_0<0$, occasionally an evolution of (\ref{eq:reg-S3-IIA}) results instead in the singularity (\ref{eq:3O4}). As usual we also tried a numerical evolution from random values of the functions at an internal $y$. With this method, we found solutions that have (\ref{eq:3O4}) at $y_+$ and an O8 at $y_-$; or solutions with (smeared) O2s on both sides.

\subsubsection{IIB solutions}
\begin{figure}[ht]
	\centering
		\includegraphics[width=10cm]{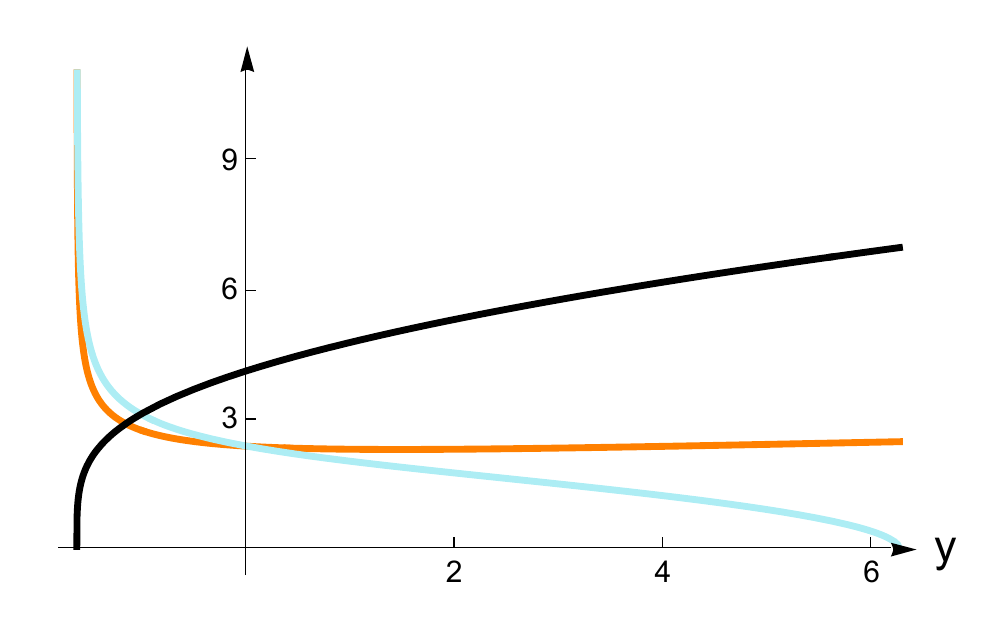}
	\caption{\small A numerical solution with M$_6$ an $\mathrm{S}^3$-fibration over a three-manifold $\Sigma_3$, interpolating between an O5-plane on the left, and a regular point on the right. The functions are $e^A$ (orange), $e^B$ (aqua), $e^C$ (black). The parameters in (\ref{eq:reg-S3-IIB}) in this case are $a_0=2.5$, $c_0=7$, $R_0=6$, $m=-1$.}
	\label{fig:S3IIB-O5-reg}
\end{figure}

Applying the S$^3\times\Sigma_3$ fibration ansatz to the IIB class of section \ref{sec:IIBbetagen} we find that the torsion classes are fixed in a similar (but not the same) fashion as for the IIA class, namely
\begin{align}
T_2&=0,~~~~~S=\frac{e^{2A-2B-C}h^{-1}(R_0e^{2C}-2e^{2B})}{8\cos\beta}\bigg[-3e^{3C}e^1\wedge e^2\wedge e^3-\frac{1}{2}e^{2B+C}\epsilon_{ijk}\mu^i\wedge \mu^j\wedge e^k\nn\\[2mm]
&+i\big(3e^{3B}\mu_1\wedge \mu_2\wedge \mu_3+\frac{1}{2}e^{B+2C}\epsilon_{ijk}\mu^i\wedge e^j\wedge e^k\big)\bigg].
\end{align}
 In addition to this we get the following ODEs for the functions of the ansatz:
\begin{align}
\partial_y\left(\frac{e^{-A+3C}h^{-1}}{\cos\beta}\right)&=\frac{R_0}{2}\partial_y\left(\frac{e^{-A+2B+C}h^{-1}}{\cos\beta}\right),~~~~
\partial_y\left(\frac{e^{-A+2B+C}h^{-1}}{\cos\beta}\right)=\frac{1}{m}\frac{e^{B+C}h'}{h^2\cos^2\beta \sqrt{1- e^{-4A}h^2}},\nn\\[2mm]
\partial_y\left(\frac{e^{2B+2C}}{\cos^2\beta}\right)&= \frac{1}{6m}\frac{e^{-3A+B}(e^{2B}R_0+6 e^{2C})hh'}{\cos\beta  \sqrt{1-e^{-4A}h^2}},~~~~\partial_y\left(\frac{e^{3B+3C}\sin\beta}{ \cos^3\beta}\right)=0.
\end{align}
These imply that
\beq
e^{3B+3C}\sin\beta= b_1  \cos^3\beta,\qquad e^{3C}-\frac{R_0}{2}e^{2B+C}=b_2e^{A}h\cos\beta
\eeq
where $b_{1,2}$ are arbitrary integration constants.\\
~~\\
We work in the gauge $h=y$. As in section \ref{ssub:S3-IIA}, we have been able to find a local analytic solution where the S$^3$ shrinks smoothly at $y=a_0^2$:
\begin{align}
	\nonumber
	e^A&= a_0+ \frac1{8a_0}(y-a_0^2) + O(a_0^2-y)^2 \, ,\qquad
	e^B = -\frac1m\sqrt{\frac23 (a_0^2-y)} + \frac{4 R_0 a_0^2 - 8 m^2 c_0^2}{108\sqrt{6} m^3}O(a_0^2-y)^{3/2}\,\\
	\label{eq:reg-S3-IIB} 
	e^C &= c_0 +\frac{8 a_0^2 R_0 - 27 m^2 c_0^2}{72 m^2 a_0^2 c_0}(a_0^2-y) + O(a_0^2-y)^2\,.
\end{align}
This is almost identical to (\ref{eq:reg-S3-IIA}) at this level of approximation, but differs more markedly from it at higher orders. Evolving it numerically, the solution stops at a singularity with local metric
\begin{equation}\label{eq:O5}
	ds^2 \sim (y-y_0)^{-1/2}\left(ds^2(\text{AdS}_3) + ds^2(\text{S}^3)\right) + (y-y_0)^{1/2}(ds^2(\Sigma_3) +dy^2)\quad [\text{(smeared) O5}]\,.
\end{equation}
When $\Sigma_3$ is hyperbolic, we encounter the same problem we described below (\ref{eq:smO4}): the cone $C(\Sigma_3)$ is not physically sensible, so we cannot consider (\ref{eq:O5}) as an O5 placed at its tip. Rather, we need to fall back on the alternative interpretation of an O5 smeared along $\Sigma$. But when $\Sigma=$S$^3$, (\ref{eq:O5}) should represent a genuine O5-plane singularity, without any smearing. Both cases are possible numerically.\\ 
~~\\
So in particular in this class we are able to find solutions whose only singularity consists of a fully localized O5-plane. We show an example in Fig.~\ref{fig:S3IIB-O5-reg}.

A less interesting type of numerical solution we found has (\ref{eq:O5}) at $y=y_-$, and the singularity with four smeared O$p$-planes (\ref{eq:4O}) at $y=y_+$.

\subsection{Summary of physical solutions} 
\label{sub:summary}

We found several types of concrete solutions in this section. We list here only those that have no smeared O-planes.

\begin{itemize}
	\item In IIA, the solution in Fig.~\ref{fig:nkIIA-O8-reg} has M$_7=$  a fibration of a nearly K\"ahler manifold M$_6$ over an interval $[y_-,y_+]$; it has an O8-plane at $y=y_-$, and a conical G$_2$-singularity at $y=y_+$. When M$_6=\mathrm{S}^6$, the latter is a regular point. 
	\item In IIA, the solution in Fig.~\ref{fig:S2IIA-O8-reg} has now an M$_6=\mathrm{S}^2$ fibration over an Einstein self-dual M$_4$, again all fibred over an interval. As in the previous case, there is an O8 at $y=y_-$, but $y=y_+$ is always fully regular.
	\item In IIA, for the analytic solution (\ref{eq:S3-analytic}) the internal space is an $\mathrm{S}^3$ fibration over a maximally symmetric $\Sigma_3$, all fibred over an interval. Again there is an O8 at $y=y_-$, and a regular point at $y=y_+$. This is interpreted as a compactification on $\Sigma_3$ of the AdS$_6$ solution in \cite{Brandhuber:1999np}. However, we also found a two-parameter numerical generalization, shown in in Fig.~\ref{fig:S3IIA-O8-reg}. 
	\item In IIB, again with M$_6=$ an S$^3$-fibration over $\Sigma_3$, fibred over an interval, the solution in Fig.~\ref{fig:S3IIB-O5-reg} has an O5 at $y=y_-$, and is regular at $y=y_+$. 
\end{itemize}

We should stress again that our search was far from exhaustive. We picked only some simple possibilities for M$_6$; it is very likely that many other solutions exist. Even for the ans\"atze we did try, it would have been possible to look for more general solutions, for example including D8-branes and/or O8-planes with a finite dilaton (as opposed to (\ref{eq:O8}), where it diverges as $e^\phi \sim |y-y_-|^{-5/4}$). On the other hand, a more complete treatment would now be required to make sure that the supergravity approximation is under control, and that flux quantization can be imposed. For the types of solutions under consideration in this paper, both of these are expected to work easily, in part because of various rescaling symmetries of the supergravity equations of motion. We do not carry out this explicitly in this paper, but it would be needed to analyze the correspondence with CFT$_2$'s.


\section*{Acknowledgments}
NM would like to thank C.~Couzens, A.~Legramandi and A.~Passias for collaboration on related projects; we would like to thank E.~Witten for interesting correspondence. NM is supported by the Spanish government grant PGC2018-096894-B100. AT is supported in part by INFN and by MIUR-PRIN contract 2017CC72MK003.

\appendix

\section{AdS$_3$ bi-linears}\label{sec:app1}
In this appendix we give details of the bi-linears following from the two Killing spinors on AdS$_3$ $\zeta_{\pm}$ charged under the SL(2)$_{\pm}$ subgroup of SO(2,2). These solve the Killing spinor equation as
\beq
\nabla_{\mu}\zeta_{\pm}=\pm \frac{m}{2}\gamma_{\mu} \zeta_{\pm}
\eeq
and we choose real gamma matrices, such that $\zeta_{\pm}$ can also be taken to be real we also take them to be unit norm without loss of generality. Using the Killing spinor equation, or via direct computation with respect to specific gamma matrices and  specific solutions to the Killing spinor equation, one can establish that
\beq
\zeta_{\pm}\otimes \overline{\zeta}_{\pm}= \frac{1}{2}\big(v^{\mp}\mp f^{-1} u\wedge v^{\mp}\big),\qquad\zeta_{\pm}\otimes\overline{\zeta}_{\mp}= \frac{1}{2}\big( \pm f-u+\frac{1}{2} f^{-1}v^+\wedge v^-\mp f \text{vol}(\text{AdS}_3) \big)
\eeq
where $(v^{\pm},u)$ are  one-forms obeying the conditions 
\begin{align}
dv^{\pm}&=2 m f^{-1}v^{\pm}\wedge u,& df&=-m u,\nn\\[2mm]
u\lrcorner\text{vol}(\text{AdS}_3)&=\frac{1}{2}f^{-1}v^+\wedge v^-,& v^{\pm}\lrcorner\text{vol}(\text{AdS}_3)&=\pm f^{-1}v^{\pm}\wedge u.
\end{align}
(The $\pm$ is just a label.) Further one can show that
\beq
\nabla_{(\mu}v^{\pm}_{\nu)}=0,\qquad\nabla_{(\mu}u_{\nu)}=-m f g(\text{AdS}_3)_{\mu\nu},
\eeq
so that $(v^{\pm})^{\mu}\partial_{\mu}$ are both Killing vectors while $(u)^{\mu}\partial_{\mu}$ is only a conformal Killing vector.  They obey the following interior product relations
\beq
v^{\pm}\lrcorner v^{\pm}=v^{\pm}\lrcorner u=0,\qquad v^{\pm}\lrcorner  v^{\mp}=-2u\lrcorner u =-2f^2,
\eeq
so $v^{\pm}$ are null and orthogonal to $u$, such that any linear combination of $v^{\pm}$ with positive coefficients is strictly time-like. This will eventually lead to a time-like $d=10$ Killing vector in the next section.\\
~~\\
A particular parameterisation of AdS$_3$ is
\beq
ds^2(\text{AdS}_3)= e^{2m r}(-dt^2+ dx^2)+dr^2.
\eeq
In terms of this, the Killing spinors that are Poincar\'e invariant in $(t,x)$ are
\beq
\zeta_+= e^{\frac{m}{2}r}\left(\begin{array}{c} 1\\0\end{array}\right),\qquad \zeta_-=e^{\frac{m}{2}r}\left(\begin{array}{c} 0\\1\end{array}\right),
\eeq
when we take $\gamma_{\mu}=(i \sigma_2,\sigma_1,\sigma_3)_{\mu}$. One then has
\beq
f= e^{m r},\qquad v^{\pm}=e^{2m r}(dt\pm dx),\qquad u=- e^{mr} dr.
\eeq
Using a specific parameterisation such as this is probably the easiest way to derive the identities in this section.

\section{General ${\cal N}=(1,0)$ conditions}
In this appendix, for the first time,  we present all the ${\cal N}=(1,0)$ conditions for AdS$_3$ without assumption. We make use of an existing classification of generic supersymmetric solutions in type II supergravities presented in  \cite{Tomasiello:2011eb}.\\
~~\\
The $d=10$ Majorana--Weyl Killing spinors for ${\cal N}=(1,0)$ supersymmetric AdS$_3$ decompose as
\beq
\epsilon_1= \zeta_+\otimes \theta_+\otimes \chi^1,\qquad \epsilon_2= \zeta_+\otimes \theta_{\mp}\otimes \chi^2
\eeq
where $\pm$ in $\theta_{\pm}$ is labeling 10d chirality while on $\zeta_{+}$ (the AdS$_3$ Killing spinor defined in appendix \ref{sec:app1}) it labels the SL(2)$_+$ subgroup of SO$(2,2)=$SL(2)$_+\times$ SL(2)$_-$. The bosonic fields decompose as in \eqref{eq:gendecomp} so as to respect the SO(2,2) isometry of AdS$_3$. Following \cite{Tomasiello:2011eb} one can define the following one-form bi-linears
\beq
K= \frac{1}{64}(\overline{\epsilon}_1\Gamma_{M}\epsilon_1+\overline{\epsilon}_2\Gamma_{M}\epsilon_2)dX^M,\qquad\tilde{K}= \frac{1}{64}(\overline{\epsilon}_1\Gamma_{M}\epsilon_1-\overline{\epsilon}_2\Gamma_{M}\epsilon_2)dX^M\label{eq:KKtildeappendixdef}
\eeq
where $K^{M}$ are necessarily the components of a 10 dimensional Killing vector that is either time-like or null.  Decomposing the 10 dimensional gamma matrices as
\beq
\Gamma_{\mu}= e^{A} \gamma_{\mu}\otimes \sigma_3\otimes \mathbb{I},\qquad\Gamma_a= \mathbb{I}\otimes \sigma_1\otimes \gamma_a,\qquad i\gamma_{1234567}=1
\eeq
implies that the 10d intertwiner defining Majorana conjugation as $\epsilon^c= B^{(10)}\epsilon^*$ and chirality matrix $\hat\Gamma$ are respectively
\beq
B= \mathbb{I}\otimes \sigma_3\otimes B,\qquad\hat\Gamma=\mathbb{I}\otimes\sigma_2\otimes \mathbb{I}, 
\eeq
where $B\gamma_a B^{-1}=-\gamma_a^{*}$, $B B^*=\mathbb{I}$. As such we should take
\beq
\theta_+=\frac{1}{\sqrt{2}}\left(\begin{array}{c}1\\-i\end{array}\right),\qquad\theta_-=\frac{1}{\sqrt{2}}\left(\begin{array}{c}1\\i\end{array}\right),
\eeq
$\zeta_{+}$ to be real and $(\chi^{1,2})^c= B (\chi^{1,2}_{\pm})^*=(\chi^{1,2})$ so that $\epsilon_{1,2}$ are Majorana--Weyl as required. Using this we can now refine \eqref{eq:KKtildeappendixdef} as
\begin{align}
K=\frac{e^{A}}{64}(|\chi^1_-|^2+|\chi^2_-|^2)v^+,~~~~~
\tilde{K}&=\frac{e^{A}}{64}(|\chi^1_-|^2-|\chi^2_-|^2)v^+,
\end{align}
making $K$ null. The first conditions for supersymmetry are that $K^M\partial_M$ should be Killing, and that NS the three-form $H$ and the one-form $\tilde K_M$ obey 
\beq\label{eq:NSfluxcond}
d\tilde K= K\lrcorner H.
\eeq
These impose that
\beq\label{eq: neq1norms}
|\chi^1|^2\pm |\chi^2|^2=c_{\pm} e^{\pm A},~~~~~ c_+ e^{3A}h_0= -2m c_- 
\eeq
where $c_+$, $c_-$ are constants, the former strictly positive. Another set of necessary conditions is given by
\beq\label{eq:10dbilinearcond}
d_H(e^{-\Phi}\Psi^{(10)})=- (K\lrcorner+\tilde{K}\wedge)F_{10},~~~~ \Psi^{(10)}=\epsilon_1\otimes \overline{\epsilon}_2.
\eeq 
It is not hard to show that the bi-linear is given by
\begin{align}
4\Psi^{(10)}&= \mp e^A v^-\wedge \Psi_{\mp}- f^{-1} e^{2A}u\wedge v^-\wedge\Psi_{\pm},~~~~ \chi_1\otimes \chi_2^{\dag}= \Psi_++ i \Psi_-,
\end{align}
while the term involving the flux becomes
\begin{align}
(K\lrcorner+\tilde{K}\wedge)F_{10}=\frac{1}{32}\bigg[\frac{c_-}{2}v^-\wedge f_{\pm} -\frac{c_+}{2}f^{-1}e^{3A}v^-\wedge u\wedge \star_7\lambda(f_{\pm})\bigg].
\end{align}
Given this, the 10d bi-linear constraints reduce to the 7d ones
\beq\label{eq:bispinorcondneq1}
d_{H_3}(e^{A-\Phi}\Psi_{\mp})\pm \frac{c_-}{16}f_{\pm}=0,~~~~d_{H_3}(e^{2A-\Phi}\Psi_{\pm})\mp 2m e^{A-\Phi}\Psi_{\mp}=\frac{c_+}{16}e^{3A}\star_7\lambda(f_{\pm}).
\eeq
The final necessary conditions are given by (3.1c) and (3.1d) of \cite{Tomasiello:2011eb}. Dealing with these is a much more lengthy computation, so from here we shall only sketch the derivations giving some key intermediate results. Given the AdS$_3$ ansatz, (3.1c) for instance simplifies considerably, reducing to
\beq\label{eq:10dpairing1}
\bigg(e_{1+}\Psi^{(10)}e_{2+},~\Gamma^{MN}\bigg[dA\wedge e_{2+} \Psi^{(10)}-\frac{1}{2}\big(e^{-A}m+\frac{1}{2}h_0\big)Pe_{2+}\Psi^{(10)}-2 e^{\Phi}f_{\pm} \bigg]\bigg)=0
\eeq
where 
\beq
e_{a+}=- \frac{8 e^{A}}{f^2|\chi_a|^2} v^+,~~~~~ P =\mathbb{I}\otimes \sigma_3\otimes \mathbb{I},
\eeq
and the bracket is $(\alpha, \beta) \equiv (\alpha \wedge \lambda(\beta))_{10}$. In principle \eqref{eq:10dpairing1} could give 3 independent conditions for $(M,N)$, i.e.~those aligned along AdS$_3$, M$_7$ or mixed directions respectively. However, due to $v^+. v^+=0$, when $(M,N)=(a,b)$ this condition is trivial. With some effort one can show that $(M,N)=(\mu,a)$ is equivalent to 
\beq
\big(\Psi_{\pm},~\gamma_a(e^{-\Phi} dA\wedge\Psi_{\mp}\pm \frac{|\chi_2|^2}{8}f_{\pm})\big)_7=0,
\eeq
where the bracket is now  $(X,Y)_7=X\wedge \lambda(Y)\big\lvert_7$. It is then possible to show that this is implied by \eqref{eq:bispinorcondneq1} using  properties of the pairing. Thus the only new conditions follow from the AdS$_3$ directions --- one can show that this holds true for (3.1d) of \cite{Tomasiello:2011eb} also. It turns that (3.1c) and (3.1d) of \cite{Tomasiello:2011eb} yield conditions depending on $|\chi_2|^2$ and $|\chi_1|^2$ respectively, which given \eqref{eq: neq1norms} are not actually independent; they are equivalent to a single condition
\beq
(\Psi_{\mp},f_{\pm})_7= \mp \frac{c_+ m+ \frac{1}{2}e^{-A} c_- h_0}{4} e^{-\Phi}\text{vol}(\text{M}_7).
\eeq
In summary, in conventions where $i\gamma_{1234567}=1$, the necessary conditions for ${\cal N}=(1,0)$ supersymmetry in full generality are
\begin{align}
&d_{H_3}(e^{A-\Phi}\Psi_{\mp})\pm \frac{c_-}{16}f_{\pm}=0,~~~~d_{H_3}(e^{2A-\Phi}\Psi_{\pm})\mp 2m e^{A-\Phi}\Psi_{\mp}=\frac{c_+}{16}e^{3A}\star_7\lambda(f_{\pm}),\\[2mm]
&(\Psi_{\mp},f_{\pm})_7= \mp \frac{c_+ m+ \frac{1}{2}e^{-A} c_- h_0}{4} e^{-\Phi}\text{vol}(\text{M}_7),~~~~|\chi^1|^2\pm |\chi^2|^2=c_{\pm} e^{\pm A},~~~~~  c_+e^{3A}h_0= -2m c_-, \nn
\end{align}
where one can in fact fix $c_+$ to any value one chooses without loss of generality. Notice that for $(c_+=2,c_-=0)$ we reproduce the conditions presented in  \cite{Passias:2020ubv} which shares our conventions. The conditions for ${\cal N}=(0,1)$ are quite similar, they are  given by sending $m \to - m$ everywhere it appears in the above expressions.

\section{Derivation of supersymmetry conditions for ${\cal N}=(1,1)$ AdS$_3$ }\label{sec:derivationsusy}
In this appendix we give a detailed derivation of the supersymmetry conditions for ${\cal N}=(1,1)$ AdS$_3$ summarised in section \ref{sec:bispingen}. We again make use of  \cite{Tomasiello:2011eb} and use the same conventions as the preceding appendices.\\
~\\
The $d=10$ Majorana--Weyl Killing spinors for ${\cal N}=(1,1)$ supersymmetric AdS$_3$ decompose as
\beq
\epsilon_1= \zeta_+\otimes \theta_+\otimes \chi^1_++\zeta_-\otimes \theta_+\otimes \chi^1_-,\qquad \epsilon_2= \zeta_+\otimes \theta_{\mp}\otimes \chi^2_++\zeta_-\otimes \theta_{\mp}\otimes \chi^2_-
\eeq
where $\pm$ in $\theta_{\pm}$ is labeling 10d chirality (so the upper/lower signs should be taken in IIA/IIB), while on $\zeta_{\pm}$ and $\chi^{1,2}_{\pm}$ it is just a label. Specifically  $\zeta_{\pm}$ are two independent AdS$_3$ Killing spinors defined in appendix \ref{sec:app1}. Plugging these spinors into \eqref{eq:KKtildeappendixdef} we find the 10d 1-forms
\begin{align}
K=&\frac{1}{32}\bigg(\frac{e^{A}}{2}(|\chi^1_-|^2+|\chi^2_-|^2)v^++\frac{e^{A}}{2}(|\chi^1_+|^2+|\chi^2_+|^2)v^-- e^{A}(\chi^{1\dag}_-\chi^1_++\chi^{2\dag}_-\chi^2_+)u- f\xi\bigg),\nn\\[2mm]
\tilde{K}&=\frac{1}{32}\bigg(\frac{e^{A}}{2}(|\chi^1_-|^2-|\chi^2_-|^2)v^++\frac{e^{A}}{2}(|\chi^1_+|^2-|\chi^2_+|^2)v^-- e^{A}(\chi^{1\dag}_-\chi^1_+-\chi^{2\dag}_-\chi^2_+)u- f\tilde{\xi}\bigg)
\end{align}
where
\beq
\xi_a=-i (\chi^{1\dag}_+\gamma_a\chi^1_-\mp \chi^{2\dag}_+\gamma_a\chi^2_-),~~~\tilde\xi_a=-i (\chi^{1\dag}_+\gamma_a\chi^1_-\pm \chi^{2\dag}_+\gamma_a\chi^2_-).
\eeq
The fact that $K^{M}\partial_M$ should be a 10d Killing vector imposes the following 7d conditions
\begin{subequations}
\begin{align}
&\nabla_{(a}\xi_{b)}=0,\label{eq:killveccond1}\\[2mm]
&{\cal L}_{\xi}A- m e^{-A}(\chi^{1\dag}_-\chi^1_++\chi^{2\dag}_-\chi^2_+)=0,\label{eq:killveccond2}\\[2mm]
&d(e^{-A}(\chi^{1\dag}_-\chi^1_++\chi^{2\dag}_-\chi^2_+))-m e^{-2A}\xi=0,\label{eq:killveccond4}\\[2mm]
&d(e^{-A}(|\chi^1_+|^2+|\chi^2_+|^2))=d(e^{-A}(|\chi^1_-|^2+|\chi^2_-|^2))=0,\label{eq:killveccond3}
\end{align}
\end{subequations}
 from which it is clear that for a generic solution $\xi^a\partial_a$ is a Killing vector of the internal space but not the warp factor --- unless we have $\xi=0$. Another necessary condition is that the NS flux obeys \eqref{eq:NSfluxcond} which leads to
\begin{align}
&d\tilde\xi=\xi\lrcorner H_3,\nonumber\\[2mm]
&h_0(\chi^{1\dag}_-\chi^1_++\chi^{2\dag}_-\chi^2_+)=0,\nn\\[2mm]
&d(e^{A}(\chi^{1\dag}_-\chi^1_+-\chi^{2\dag}_-\chi^2_+))+m \tilde{\xi}=0,\\[2mm]
&d(e^{A} (|\chi^1_+|^2-|\chi^2_+|^2))=d(e^{A} (|\chi^1_-|^2-|\chi^2_-|^2))=0,\nn\\[2mm]
&2m (|\chi^1_-|^2-|\chi^2_-|^2)-e^{A}h_0 (|\chi^1_-|^2+|\chi^2_-|^2)=2m (|\chi^1_+|^2-|\chi^2_+|^2)+e^{A}h (|\chi^1_+|^2+|\chi^2_+|^2)=0\nn.
\end{align}
So we must fix either
\beq
h_0=0,~~~~\text{or}~~~~\chi^{1\dag}_-\chi^1_++\chi^{2\dag}_-\chi^2_+=0,
\eeq
or both. When $\xi \neq 0$, (\ref{eq:killveccond4}) implies we should fix $h_0=0$. In fact it is fairly easy to show that in this case the AdS$_3$ factor we are assuming gets enhanced to AdS$_4$ in both the metric and the fluxes (at least locally). To see this one can solve \eqref{eq:killveccond4} locally as 
\beq
e^{-A}(\chi^{1\dag}_-\chi^1_++\chi^{2\dag}_-\chi^2_+)= \rho(r), \qquad m \xi=e^{2A} \rho' dr.
\eeq
$r$ now plays the role of a local coordinate and $\rho'$ parameterises diffeomorphism invariance in this direction. We can implicitly fix this invariance by taking
\beq
\xi^a\partial_{a}= \partial_{r}\qquad\Rightarrow\qquad ||\xi||^2=\frac{1}{m} e^{2A} \rho' .
\eeq
$\xi$ and the vectors orthogonal to it define two distributions, which are integrable by (\ref{eq:killveccond4}); in other words $\xi$ is hypersurface orthogonal. There thus exist coordinates such 
\beq
ds^2(\text{M}_7)= (e^{\xi})^2+ds^2(\text{M}_6), \qquad e^{\xi}=\frac{\xi}{||\xi||}= e^{A}\sqrt{\frac {\rho'}m} dr.
\eeq
Now $\xi$ being a Killing vector implies
\beq
e^{A}= \sqrt{\frac m{\rho'}} e^{A_4},\qquad \partial_r A_4=0.
\eeq
Plugging this into \eqref{eq:killveccond2} then leads without loss of generality to
\beq
\rho= \frac{1}{m}\tanh r,
\eeq
and the 10d metric becomes
\beq
ds^2= e^{2A_4}\bigg[m^2\cosh^2 r ds^2(\text{AdS}_3)+ dr^2\bigg]+ds^2(\text{M}_6)
\eeq
which is warped AdS$_4$, not warped AdS$_3$. This pattern persists with the fluxes as well, at least away from the loci of possible sources, so in regular regions of a solution
\beq
\xi \neq 0\qquad\Rightarrow \qquad\text{AdS}_3 \to \text{AdS}_4,
\eeq
so at best solutions with $\xi \neq 0$ can generalise AdS$_4$ to cases with sources placed along the AdS radial direction, and indeed we have not established that  even  this is necessarily possible.\\
~~\\
Thus for AdS$_3$ vacua we can  fix $\xi=0$ without loss of generality. This truncates the conditions derived thus far to:
\begin{subequations}\label{eq:thusfar}
\begin{align}
&\xi=0,\label{eq:xicond}\\[2mm]
&\chi^{1\dag}_-\chi^1_++\chi^{2\dag}_-\chi^2_+=0,\label{eq:pluscond}\\[2mm]
&d(e^{A}(\chi^{1\dag}_-\chi^1_+-\chi^{2\dag}_-\chi^2_+))+m \tilde{\xi}=0,\label{eq:xicondtilde}\\[2mm]
&|\chi^1_+|^2\pm|\chi^2_+|^2 = e^{\pm A}c_{\pm}^+, \qquad |\chi^1_-|^2\pm|\chi^2_-|^2 = e^{\pm A}c_{\pm}^-,\label{eq:normsconds}\\[2mm]
&2m c^+_{-}+ e^{3A}h_0 c^+_{+} = 0,\qquad 2m c^-_{-}- e^{3A}h_0 c^-_{+} = 0\label{eq:hconds},
\end{align}
\end{subequations}
which imply $c^+_{-}c^-_{+}+ c^-_{-}c^+_{+}=0$. With these restrictions $K$ becomes necessarily time-like for ${\cal N}=(1,1)$. It also follows  that 
\beq \label{eq:gneq0}
\chi^{1\dag}_-\chi^1_+-\chi^{2\dag}_-\chi^2_+ \neq 0,
\eeq
or rather more specifically that  $\chi^{1\dag}_-\chi^1_+-\chi^{2\dag}_-\chi^2_+$ cannot be set to zero everywhere. 
Indeed from \eqref{eq:xicond}, \eqref{eq:pluscond}, \eqref{eq:xicondtilde} and the definitions of $(\xi,\tilde{\xi})$ it follows that
\beq \label{eq:xx-xx=0}
\chi^{1\dag}_-\chi^1_+-\chi^{2\dag}_-\chi^2_+= 0\quad\Rightarrow \quad \chi^{1\dag}_-\chi^1_+=\chi^{2\dag}_-\chi^2_+= \chi^{1\dag}_-\gamma_a\chi^1_+= \chi^{2\dag}_-\gamma_a\chi^2_+=0,
\eeq
where $\{\chi, \gamma_a \chi\}$ is a basis for the space of spinors, for any $\chi$. Applying this to (\ref{eq:xx-xx=0}) implies that one of $\chi^1_{\pm}$ would need to be zero, and similarly for $\chi^2_\pm$. So when $\chi^{1\dag}_-\chi^1_+-\chi^{2\dag}_-\chi^2_+= 0$ everywhere only ${\cal N}=1$ supersymmetry can be realised.
~\\
We now turn our attention to the conditions \eqref{eq:10dbilinearcond}: one can show the 10d bi-linear decompose as
\begin{align}
4\Psi^{(10)}&=  f(\Psi_{\pm}^{+-}-\Psi_{\pm}^{-+})\mp e^A\big( v^-\wedge \Psi^{++}_{\mp}+ v^+\wedge \Psi^{--}_{\mp}-u\wedge (\Psi^{+-}_{\mp}+\Psi^{-+}_{\mp})\big)\\[2mm]
&- f^{-1} e^{2A}u\wedge (v^-\wedge\Psi^{++}_{\pm}-v^+\wedge\Psi^{--}_{\pm})\nn\\[2mm]
&+\frac{1}{2} f^{-1}e^{2A}v^+\wedge v^-\wedge(\Psi^{+-}_{\pm}+\Psi^{-+}_{\pm})\pm f e^{3A}\text{vol}(\text{AdS}_3)\wedge(\Psi^{+-}_{\mp}-\Psi^{-+}_{\mp}), \nn
\end{align}
where we defined
\beq
\Psi^{st}_++ i \Psi^{st}_-= \chi^1_s\otimes \chi^{2\dag}_t,~~~ s,t=\pm.
\eeq
Likewise one can show that 
\begin{align}
(K\lrcorner+\tilde{K}\wedge)F_{10}=\frac{1}{32}\bigg[&f e^{3A}\text{vol}(\text{AdS}_3)\wedge \tilde \xi\wedge \star_7\lambda(f_{\pm})- f \tilde{\xi}\wedge f_{\pm}+ \frac{1}{2}(c^-_-v^++c^+_-v^-)\wedge f_{\pm}\\[2mm]
-&e^{A}(\chi^{1\dag}_-\chi^1_+-\chi^{2\dag}_-\chi^2_+)u\wedge f_{\pm}+ \frac{1}{2}f^{-1}e^{3A}(c^-_+v^+-c^+_+v^-)\wedge u\wedge \star_7\lambda(f_{\pm})\bigg]\nn.
\end{align}
Putting this all together we find \eqref{eq:10dbilinearcond} is equivalent to the 7d conditions
\begin{subequations}\label{eq:bispin}
\begin{align}
&d_{H_3}(e^{A-\Phi}\Psi^{++}_{\mp})\pm \frac{c^+_-}{16}f_{\pm}=0,\label{eq:bispin1}\\[2mm]
&d_{H_3}(e^{A-\Phi}\Psi^{--}_{\mp})\pm \frac{c^-_-}{16}f_{\pm}=0,\label{eq:bispin2}\\[2mm]
&d_{H_3}(e^{2A-\Phi}\Psi^{++}_{\pm})\mp 2m e^{A-\Phi}\Psi^{++}_{\mp}=\frac{c^+_+}{16}e^{3A}\star_7\lambda(f_{\pm}),\label{eq:bispin3}\\[2mm]
&d_{H_3}(e^{2A-\Phi}\Psi^{--}_{\pm})\pm 2m e^{A-\Phi}\Psi^{--}_{\mp}=\frac{c^-_+}{16}e^{3A}\star_7\lambda(f_{\pm}),\label{eq:bispin4}\\[2mm]
&d_{H_3}(e^{2A-\Phi}(\Psi^{+-}_{\pm}+\Psi^{-+}_{\pm}))=0,\label{eq:bispin5}\\[2mm]
&d_{H_3}(e^{-\Phi}(\Psi^{+-}_{\pm}-\Psi^{-+}_{\pm}))=\frac{1}{8}\tilde{\xi}\wedge f_{\pm},\label{eq:bispin6}\\[2mm]
&d_{H_3}(e^{A-\Phi}(\Psi^{+-}_{\mp}+\Psi^{-+}_{\mp}))\pm m e^{-\Phi}(\Psi^{+-}_{\pm}-\Psi^{-+}_{\pm})=\mp\frac{1}{8}e^{A}(\chi^{1\dag}_-\chi^1_+-\chi^{2\dag}_-\chi^2_+)f_{\pm},\label{eq:bispin7}\\[2mm]
&d_{H_3}(e^{3A-\Phi}(\Psi^{+-}_{\mp}-\Psi^{-+}_{\mp}))\pm e^{3A-\Phi}h_0(\Psi^{+-}_{\pm}-\Psi^{-+}_{\pm})\pm 3m e^{2A-\Phi}(\Psi^{+-}_{\pm}+\Psi^{-+}_{\pm})= \pm \frac{1}{8}e^{3A}\tilde{\xi}\wedge \star_7\lambda(f_{\pm}).\label{eq:bispin8}
\end{align}
\end{subequations}
The first four of these are simply the distinct conditions one would get from an ${\cal N}=(1,0)$ and ${\cal N}=(0,1)$ 10 dimensional bi-linear respectively (see the preivous appendix). The Bianchi identities and equations of motion of the RR flux away from the loci of sources impose that
\beq \label{eq:bianchi}
d_{H_3}f_{\pm}=0,\qquad d_{H_3}(e^{3A}\star_7\lambda(f_{\pm}))+ e^{3A}h_0 f_{\pm}=0.
\eeq
Notice that when the $dH_3=0$ is assumed and given \eqref{eq:hconds},  $d_{H_3}$\eqref{eq:bispin1}, $d_{H_3}$\eqref{eq:bispin2} and $d_{H_3}$\eqref{eq:bispin6} combined with \eqref{eq:bispin5} imply the first of these when respectively $c^+_-,c^-_-$ and $\chi^{1\dag}_-\chi^1_+-\chi^{2\dag}_-\chi^2_+$ are assumed to be non zero; the latter indeed cannot vanish for us, as we concluded in (\ref{eq:gneq0}). Similarly $d_{H_3}$\eqref{eq:bispin3} and $d_{H_3}$\eqref{eq:bispin4} reproduce the second in (\ref{eq:bianchi}); this is true in general, as both $c^{\pm}_+\neq 0$. Notice also that \eqref{eq:bispin1}--\eqref{eq:bispin2} imply that in IIA
\beq
c^{\pm}_- F_0=0.
\eeq
In other words a non trivial Romans mass $F_0\neq 0$ is only possible when $c^{\pm}_-=0$, which by (\ref{eq:hconds}) implies that the electric NS flux $h_0=0$; conversely, $h_0\neq 0$ is only possible when $F_0=0$.  Further, in IIA the zero-form part of (\ref{eq:bispin7}), (\ref{eq:bispin8}) imply
\begin{align}
&e^{3A-\Phi}h_0(\Psi^{+-}_{0}-\Psi^{-+}_{0})+ 3m e^{2A-\Phi}(\Psi^{+-}_{0}+\Psi^{-+}_{0})=0,\nn\\[2mm]
&m e^{-\Phi}(\Psi^{+-}_{0}-\Psi^{-+}_{0})=-\frac{1}{8}e^{A}(\chi^{1\dag}_-\chi^1_+-\chi^{2\dag}_-\chi^2_+)F_0.
\end{align}
The second of these defines $F_0$. Since one of $(h_0,F_0)$ is zero, we always have
\beq
\Psi^{-+}_{0}+\Psi^{+-}_{0}=0,\qquad h_0(\Psi^{-+}_{0}-\Psi^{+-}_{0})=0,~~~
\eeq
In IIB one can always use the SL(2,$\mathbb{R}$) invariance to move to a duality frame where $h_0=0$, so one can safely fix $h_0=0$ and so $c^-_{\pm}=0$  modulo S-duality.  We now note that one cannot set $\tilde{\xi}=0$ and preserve ${\cal N}=(1,1)$ simultaneously: given the rest of the constraints this would mean that all the bi-linears $(\Psi^{\pm\pm},\Psi^{\pm\mp})$ are proportional to each other, it then quickly becomes apparent that \eqref{eq:bispin4}--\eqref{eq:bispin5} cannot be made consistent with \eqref{eq:bispin7} when $\tilde{\xi}=0$.\\
~~\\
 It is a generic feature of this kind of reduction of 10 dimensional bi-linears on a warped product that some of the conditions \eqref{eq:bispin1}--\eqref{eq:bispin8} will be implied by the others.  To see this one can generally  exploit two identities that follow from the definition of $\Psi$ and from supersymmetry: 
\beq
{\cal L}_K\Psi^{(10)}=0,\qquad (K\lrcorner+\tilde{K}\wedge)\Psi^{(10)}=0
\eeq
In the case at hand, since $K$ only has components along $v^{\pm}$, it is simple to show that the first of these holds trivially, and it is the second that yields some constraints on the 7d bi-linears. One can show that this requires
\begin{subequations}
\begin{align}
&\tilde{\xi}\wedge (\Psi^{+-}_{\pm}-\Psi^{-+}_{\pm})= \pm e^{A}(c^-_+ \Psi^{++}_{\mp}+ c^+_+  \Psi^{--}_{\mp}),\label{eq:identities1}\\[2mm]
&\tilde{\xi}\wedge (\Psi^{+-}_{\mp}+\Psi^{-+}_{\mp})=\pm\big(e^A (c^-_+\Psi^{++}_{\pm}-c^+_+\Psi^{--}_{\pm})+(\chi^{1\dag}_-\chi^1_+-\chi^{2\dag}_-\chi^2_+)(\Psi^{+-}_{\pm}-\Psi^{-+}_{\pm})\big),\\[2mm]
&2 e^{A}\tilde{\xi}\wedge \Psi^{--}_{\mp}=\pm\big(c^-_+ e^{2A}(\Psi^{+-}_{\pm}+ \Psi^{-+}_{\pm})+ c^-_-(\Psi^{+-}_{\pm}-\Psi^{-+}_{\pm})\big),\\[2mm]
&2 e^{A}\tilde{\xi}\wedge \Psi^{++}_{\mp}=\pm\big(-c^+_+ e^{2A}(\Psi^{+-}_{\pm}+ \Psi^{-+}_{\pm})+c^+_-(\Psi^{+-}_{\pm}-\Psi^{-+}_{\pm})\big),\\[2mm]
&2 \tilde{\xi}\wedge \Psi^{--}_{\pm}=\mp\big(c^-_+e^{A}(\Psi^{+-}_{\mp}-\Psi^{-+}_{\mp})+e^{-A} c^-_-(\Psi^{+-}_{\mp}+\Psi^{-+}_{\mp})-2(\chi^{1\dag}_-\chi^1_+-\chi^{2\dag}_-\chi^2_+)\Psi^{--}_{\mp}\big),\\[2mm]
&2 \tilde{\xi}\wedge \Psi^{++}_{\pm}=\mp\big(c^+_+e^{A}(\Psi^{+-}_{\mp}-\Psi^{-+}_{\mp})-e^{-A} c^+_-(\Psi^{+-}_{\mp}+\Psi^{-+}_{\mp})+2(\chi^{1\dag}_-\chi^1_+-\chi^{2\dag}_-\chi^2_+)\Psi^{++}_{\mp}\big),\\[2mm]
&e^{A}\tilde\xi\wedge(\Psi^{+-}_{\pm}+\Psi^{--}_{\pm})=\mp (c^-_- \Psi^{++}_{\mp}-c^+_- \Psi^{--}_{\mp}),\\[2mm]
&e^{A}\tilde\xi\wedge(\Psi^{+-}_{\mp}-\Psi^{--}_{\mp})=\mp \big((c^-_- \Psi^{++}_{\pm}+c^+_- \Psi^{--}_{\pm})-e^{A}(\chi^{1\dag}_-\chi^1_+-\chi^{2\dag}_-\chi^2_+)(\Psi^{+-}_{\pm}+\Psi^{-+}_{\pm})\big)\label{eq:identities8}.
\end{align}
\end{subequations}
Using these identities one can establish that $\tilde{\xi}\wedge$\eqref{eq:bispin7} implies \eqref{eq:bispin6} given  \eqref{eq:bispin3},  \eqref{eq:bispin4} and
\eqref{eq:xicondtilde}. Similarly one can generate \eqref{eq:bispin8} by wedging $\tilde{\xi}$ with either of \eqref{eq:bispin3} and \eqref{eq:bispin4}, finally \eqref{eq:bispin1} and \eqref{eq:bispin2} give rise to \eqref{eq:bispin5}. Thus many of these conditions can easily be shown to be implied by others.\\
~~\\  
To have a sufficient system of conditions that imply ${\cal N}=(1,1)$ supersymmetry, strictly speaking, we should also solve  (3.1c) and (3.1d) of \cite{Tomasiello:2011eb} as we did to derive the ${\cal N}=(1,0)$ pairing constraint in the previous section. These are the hardest to deal with generically, though in certain cases they are implied such as for Mink$_d$/AdS$_d$ vacua in $d>3$. Here we will work smart rather than hard. In the previous appendix we give necessary conditions for ${\cal N}=1$ super symmetry,  here know we have sufficient conditions if we are solving two independent copies of these,  an ${\cal N}=(1,0)$ set and an ${\cal N}=(0,1)$  set -- \eqref{eq:bispin1}--\eqref{eq:bispin2} already give all the conditions from these independent ${\cal N}=1$ sub-sectors, except the pairing constraints that follow from (3.1c) and (3.1d) of \cite{Tomasiello:2011eb}.  Specifically these conditions take the form  
\begin{align}
(\Psi^{++}_{\mp},f_{\pm})_7&= \mp \frac{c^+_+ m+ \frac{1}{2}e^{-A} c^+_- h_0}{4} e^{-\Phi}\text{vol}(\text{M}_7),\nn\\[2mm]
(\Psi^{--}_{\mp},f_{\pm})_7&= \pm \frac{c^-_+ m- \frac{1}{2}e^{-A} c^-_- h_0}{4} e^{-\Phi}\text{vol}(\text{M}_7),\label{eq:pairing appendix}.
\end{align}
We have necessary conditions for supersymmetry if we also solve these -- any additional conditions that one can extract by plugging the ${\cal N}=(1,1)$ ansatz into  (3.1c) and (3.1d) of \cite{Tomasiello:2011eb} are necessarily implied by these. However, it is possible to show that the two conditions in \eqref{eq:pairing appendix} are dependent by (\ref{eq:thusfar}) and (\ref{eq:bispin}) and in particular that the linear combination
\beq
(c^-_+\Psi^{++}_{\mp}+c^+_+\Psi^{--}_{\mp},f_{\pm})_7=0,
\eeq
is implied in general. To this end one needs some identities: first off
\beq
(c^-_+\Psi^{++}_{\mp}+c^+_+\Psi^{--}_{\mp},d_{H_3}(\Psi^{+-}_{\mp}+\Psi^{-+}_{\mp}))_7=-(d_{H_3}(c^-_+\Psi^{++}_{\mp}+c^+_+\Psi^{--}_{\mp}),\Psi^{+-}_{\mp}+\Psi^{-+}_{\mp})_7.
\eeq
This follows from $(c^-_+\Psi^{++}_{\mp}+c^+_+\Psi^{--}_{\mp},\Psi^{+-}_{\mp}+\Psi^{-+}_{\mp})_6=0$ and the definition of $\lambda$. The former can be easily proved with a concrete parametrisation of the 7d bi-linears such as \eqref{eq:bispinors1}--\eqref{eq:bispinors2} (these hold for $c^+_+=c^-_+=2$ and $c_-^+=c^-_-=0$ but generalising to generic $(c_+^{\pm},~c_-^{\pm})$  is straightforward). Similarly one can show that
\begin{align}
&(c^-_+\Psi^{++}_{\mp}+c^+_+\Psi^{--}_{\mp},\Psi^{+-}_{\pm}-\Psi^{-+}_{\pm})_7=0,\nn\\[2mm]
&(c^-_+\Psi^{++}_{\mp}+c^+_+\Psi^{--}_{\mp},X_1\wedge (\Psi^{+-}_{\mp}+\Psi^{-+}_{\mp}))_7=(X_1\wedge(c^-_+\Psi^{++}_{\mp}+c^+_+\Psi^{--}_{\mp}), \Psi^{+-}_{\mp}+\Psi^{-+}_{\mp})_7=0,
\end{align}
for $X_1$ any one-form on M$_7$. Using these identities one finds that $(c^-_+\Psi^{++}_{\mp}+c^+_+\Psi^{--}_{\mp},f_{\pm})_7=0$  is indeed implied 
given \eqref{eq:bispin1}, \eqref{eq:bispin2}, \eqref{eq:bispin7}. Thus \eqref{eq:pairing appendix} is implied by just
\beq\label{eq:appendixparingsimp}
(c^-_+\Psi^{++}_{\mp}-c^+_+\Psi^{--}_{\mp},f_{\pm})_7=\mp \bigg(\frac{m c^+_+c^-_+}{2}+\frac{1}{8}e^{-A}(c^+_- c^-_+- c^-_- c^+_+)h_0\bigg) e^{-\Phi}\text{vol}(\text{M}_7).
\eeq
Establishing whether this too is redundant is  more challenging as $(c^-_+\Psi^{++}_{\mp}-c^+_+\Psi^{--}_{\mp},\Psi^{+-}_{\mp}+\Psi^{-+}_{\mp})_6\neq 0$ in general. Thus one needs to exploit the torsion classes of the SU(3)$\times$SU(3) structure that the internal manifold supports to establish whether it is also implied. In the main text we focus on the case where M$_7$ supports an SU(3)-structure; for such manifolds we find that \eqref{eq:appendixparingsimp} is indeed implied, but this remains a open question for SU(3)$\times$SU(3)-structure more broadly. \\
~\\
This concludes our analysis of ${\cal N}=(1,1)$ supersymmetric AdS$_3$; a summary of these results is given in the main text in section \ref{sec:bispingen} where we have fixed
\beq\label{eq:fixcs}
c^{+}_{+}=c^{-}_{+}=2 ~~~\Rightarrow~~~ c^{+}_{-}= -c^{-}_- = c,~~~\Rightarrow e^{3A}h_0=-m c.
\eeq
It is not hard to see that this can be achieved without loss of generality: \eqref{eq:normsconds} ensures that $\Psi^{st}$ for $s,t=\pm$ has an overall $\sqrt{c^s_+ c^t_+}$ factor in it definition. Since $c^{\pm}_-$ must also have a $c^{\pm}_{+}$ factor in their definition and $c^{\pm}_+>0$, they may be factored out of \eqref{eq:bispin1}--\eqref{eq:bispin4}. Likewise $(\chi^{1\dag}_-\chi^1_+-\chi^{2\dag}_-\chi^2_+)$ and $\tilde\xi$ contain a $\sqrt{c^+_+ c^-_+}$ factor, so $c^{\pm}_{+}$ can also be factored out of \eqref{eq:bispin5}--\eqref{eq:bispin8}, making the precise values of $c^{\pm}_+>0$ immaterial. Choosing $c^{+}_{+}=c^{-}_{+}=2$ then leads to the other conditions in \eqref{eq:fixcs} by \eqref{eq:hconds}.

\end{document}